\documentclass[11pt,a4paper]{article}
\usepackage{jheppub}
\usepackage{amssymb}   
\usepackage{amsmath}
\usepackage{graphicx}
\usepackage{caption}
\usepackage{subcaption}
\usepackage{mathrsfs}
\usepackage{url}
\usepackage{float}
\usepackage[normalem]{ulem}

\def\beq{\begin{equation}}   
\def\eeq{\end{equation}}
\def\bea{\begin{eqnarray}}  
\def\eea{\end{eqnarray}}

\def\f21{{}_2F_{1}}

\def\beq{\begin{equation}}
\def\eeq{\end{equation}}
\def\bsp#1\esp{\begin{split}#1\end{split}}

\newcommand\barparen[1]{\overset{\textbf{\fontsize{2pt}{2pt}\selectfont(--)}}{#1}}

\catcode`\@=11
\font\manfnt=manfnt
\def\Watchout{\@ifnextchar [{\W@tchout}{\W@tchout[1]}}
\def\W@tchout[#1]{{\manfnt\@tempcnta#1\relax%
  \@whilenum\@tempcnta>\z@\do{%
    \char"7F\hskip 0.3em\advance\@tempcnta\m@ne}}}
\let\foo\W@tchout
\def\dubious{\@ifnextchar[{\@dubious}{\@dubious[1]}}

\def\@dubious[#1]{%
  \setbox\@tempboxa\hbox{\@W@tchout#1}
  \@tempdima\wd\@tempboxa
  \list{}{\leftmargin\@tempdima}\item[\hbox to 0pt{\hss\@W@tchout#1}]}
\def\@W@tchout#1{\W@tchout[#1]}
\catcode`\@=12

\title{Charged Current Drell-Yan Production at N$^3$LO}

\author[a]{Claude Duhr}
\author[b]{Falko Dulat}
\author[b]{Bernhard Mistlberger}
\affiliation[a]{Theoretical Physics Department, CERN, CH-1211 Geneva 23, Switzerland.}
\affiliation[b]{SLAC National Accelerator Laboratory, 2575 Sand Hill Rd., Menlo Park, CA, 94025 USA}

\emailAdd{claude.duhr@cern.ch, falko.dulat@gmail.com, bernhard.mistlberger@gmail.com}

\preprint{CERN-TH-2020-121, SLAC-PUB-17539}

\abstract{
We present the production cross section for a lepton-neutrino pair at the Large Hadron Collider computed at next-to-next-to-next-to leading order (N$^3$LO) in QCD perturbation theory.
We compute the partonic coefficient functions of a virtual $W^{\pm}$ boson at this order.
We then use these analytic functions to study the progression of the perturbative series in different observables.
In particular, we investigate the impact of the newly obtained corrections on the inclusive production cross section of $W^{\pm}$ bosons, as well as on the ratios of the production cross sections for $W^+$, $W^-$ and/or a virtual photon.
Finally, we present N$^3$LO predictions for the charge asymmetry at the LHC.
}


\begin{document}

\maketitle


\section{Introduction}
\label{sec:introduction}

Cross sections for the production of leptons are among the ultimate precision observables measurable at the Large Hadron Collider (LHC) (see for example refs.~\cite{Aad:2016naf,CMS:2016mtd,CMS:2015ois}).
As a consequence, they provide a unique window into the inner workings of collision processes at very high energies.
The insights gained by studying them improve our understanding of the collider experiment and the fundamental mechanisms of scattering processes alike.
To derive meaningful conclusions from such observations we must put our theoretical prediction for such scattering processes at the highest possible level. 
Here, we take one significant step in this direction and  compute next-to-next-to-next-to leading order (N$^3$LO) predictions for the production cross section of a lepton-neutrino pair in QCD perturbation theory.

We focus on the charged current Drell-Yan~\cite{Drell:1970wh} (CCDY) process, where a $W$ boson is produced from the annihilation of two quarks with opposite isospin. 
In nature, the produced $W$ boson decays within the blink of an eye, but the inclusive production probability of the bosons is easily related to the probability of producing a pair of fermions consisting of a neutrino and a charged lepton with invariant mass $Q^2$.
These particles represent a stable final state and the charged lepton can be detected by the LHC experiments.

To derive theoretical predictions for the inclusive CCDY cross section we use the factorisation of hadronic cross sections into parton distribution functions (PDFs)  and partonic cross sections.
In order to achieve high precision predictions of the hadronic cross section, it is paramount to go beyond the Born approximation for the partonic cross section.
In particular, we compute the desired partonic cross sections in the framework of perturbative QCD through N$^3$LO in the perturbative expansion in the strong coupling constant.
The analytic formul\ae\ for the partonic cross sections are some of the main results of this paper and are provided in electronic form together with its arXiv submission.

We then move on and study the impact of N$^3$LO corrections on various inclusive cross sections involving vector bosons. We focus on the progression of the perturbative series in QCD for some of the cleanest hadron collider observables. When combined with the available N$^3$LO results in the literature~\cite{Anastasiou:2016cez,Mistlberger:2018etf,Banfi2016,Duhr:2019kwi,Duhr:2020kzd,Duhr:2020seh,Chen:2018pzu,Dulat:2018bfe,Dreyer:2018qbw,Dreyer:2016oyx}, our results are an important input to gauge the relevance and the impact of N$^3$LO corrections on more differential observables, like fiducial cross sections. We also study cross section ratios for vector boson production, and observe a remarkable perturbative stability for these ratios.

This article is structured as follows: In Section~\ref{sec:setup} we briefly review the computation of the N$^3$LO corrections to off-shell $W$ production. In Section~\ref{sec:pheno} we present our main result, namely phenomenological predictions for the $W$ cross section at N$^3$LO in QCD, and we discuss the main QCD uncertainties which affect the cross section at this order. In Section~\ref{sec:ratios} we extend this analysis to ratios of vector boson cross sections and the charge asymmetry at the LHC. In Section~\ref{sec:conclusion} we draw our conclusions.

\section{Setup of the computation}
\label{sec:setup}

In this paper we compute higher-order corrections in the strong coupling constant to the charged-current Drell-Yan (CCDY) cross section, i.e., the inclusive cross section for the production of a lepton-neutrino pair of invariant mass $Q^2$ at a proton-collider with center-of-mass energy $\sqrt{S}$. 
Since we are only interested in QCD corrections, the lepton-neutrino pair can only be produced via the intermediate of an (off-shell) $W$ boson.
We can then factorise the production of the $W$ boson from its subsequent decay and cast the cross section in the following form:
\beq
\label{eq:hadrdef}
Q^2 \frac{d\sigma}{dQ^2}(p\,p\to W^{\pm}\to \ell^{\pm}\,\barparen{\nu}_{\ell}) = \frac{m_W^2}{12\pi^2\,v^2}\,\frac{Q^4}{(Q^2-m_W^2)^2+m_W^2\Gamma_W^2}\,\sigma_{W^{\pm}}(\tau)\,,
\eeq
where $v$ is the vacuum expectation value and $m_W$ and $\Gamma_W$ are the mass and width of the $W$ boson, and $\sigma_{W^{\pm}}$ denotes the inclusive cross section for the production of an off-shell $W^{\pm}$ boson with virtuality $Q^2$. 
Using QCD factorisation, this cross section can be written in the form
\beq\label{eq:qcd_fac}
\sigma_{W^{\pm}}(\tau)=\tau\sigma_0 \sum_{i,j} f_i(\tau,\mu_F^2) \otimes \eta^{\pm}_{ij}(\tau,\mu_F^2,\mu_R^2) \otimes f_j(\tau,\mu_F^2)\,,
\eeq
where $\mu_F$ and $\mu_R$ denote the factorisation and renormalisation scales, and the $f_i(x,\mu_F^2)$ denote the parton density functions (PDFs) to find a parton species $i$ with momentum fraction $x$ inside the proton.
Furthermore, $\tau=\frac{Q^2}{S}$ and $\eta_{ij}^{\pm}$ is the partonic coefficient for the production of an off-shell $W$ boson from the parton species $i$ and $j$.
In the above equation we made use of convolutions defined by
\beq
f(x)\otimes g(x)=\int_x^1 \frac{dx^\prime}{x^\prime} f(x^\prime) g\left(\frac{x}{x^\prime}\right)\,,
\eeq
and we introduced the normalisation factor 
\beq
\sigma_0=\frac{ m_W^2 \pi}{  n_c Q^2 v^2},
\eeq
where $n_c$ corresponds to the number of colours in QCD. 

The partonic coefficients can be expanded into a perturbative series in the renormalised strong coupling constant $a_S = \alpha_S(\mu_R)/\pi$
\beq
\eta_{ij}^{\pm}(z) = \eta_{ij}^{\pm(0)}(z) + a_S\,\eta_{ij}^{\pm(1)}(z) + a_S^2\,\eta_{ij}^{\pm(2)}(z)+ a_S^3\,\eta_{ij}^{\pm(3)}(z)+\ldots\,.
\eeq
Above we have suppressed arguments of the functions indicating the dependence of the partonic coefficients on the perturbative scales.
At leading order (LO) in $\alpha_S$, it is only possible to produce a $W$ boson from the annihilation of two (massless) quarks with opposite isospin:
\beq
\eta_{u_i\bar{d}_j}^{+(0)}(z) = \,|V_{u_id_j}|^2\,\delta(1-z) \textrm{~~~and~~~} 
\eta_{d_i\bar{u}_j}^{-(0)}(z) = \,|V_{u_jd_i}|^2\,\delta(1-z)\,,
\eeq
where $V_{u_id_j}$ denotes the Cabibbo-Kobayashi-Maskawa quark-mixing-matrix. 
Beyond LO~\cite{Drell:1970wh}, also other partonic channels open up. 
Perturbative corrections to the CCDY cross sections have been computed at next-to-leading order (NLO) in refs.~\cite{Altarelli:1979ub,KubarAndre:1978uy,Harada:1979bj,Aurenche:1980tp} and at next-to-next-to-leading order (NNLO) in refs.~\cite{Hamberg:1990np,Anastasiou:2003ds,Melnikov:2006di,Harlander:2002wh}. 
The main result of this paper is to present for the first time phenomenological results for the production of an off-shell $W$ boson at next-to-next-to-next-to-leading oder (N$^3$LO) in perturbative QCD.
Before we discuss our results in the next sections, we review in this section the main steps of the computation of the third-order corrections to the partonic coefficients.

The partonic coefficients are computable from Feynman diagrams, with $z = Q^2/\hat{s}$ and $\hat{s}$ the partonic center-of-mass energy. 
We have followed the same strategy as that for the computation of the inclusive cross section for Higgs boson production through gluon fusion and bottom-quark fusion~\cite{Anastasiou:2015ema,Mistlberger:2018etf,Duhr:2019kwi,Duhr:2020kzd} and the inclusive Drell-Yan cross section~\cite{Duhr:2020seh}.
In particular, the results were obtained using the framework of reverse unitarity~\cite{Anastasiou2002,Anastasiou2003,Anastasiou:2002qz,Anastasiou:2003yy,Anastasiou2004a} in order to compute all required interferences of real and virtual amplitudes contributing to the N$^3$LO cross section. 
The required phase-space and loop integrals were carried out implicitly using integration-by-parts (IBP) identities~\cite{Tkachov1981,Chetyrkin1981,Laporta:2001dd},  together with the method of differential equations~\cite{Kotikov:1990kg,Kotikov:1991hm,Kotikov:1991pm,Henn:2013pwa,Gehrmann:1999as}. 
This method allows one to represent the required integrated and interfered amplitudes in terms of linear combinations of \emph{master integrals}.
The purely virtual corrections are essentially identical to the case of the production of an off-shell photon, apart from the colour structure involving a cubic Casimir operator, which is absent here because of the non-diagonal flavour-structure of the charged-current interactions. 
The three-loop corrections for virtual photon production were first computed in refs.~\cite{Gehrmann:2006wg,Heinrich:2007at,Heinrich:2009be,Lee:2010cga,Baikov:2009bg,Gehrmann:2010ue,Gehrmann:2010tu}, and  recomputed and confirmed in ref.~\cite{Duhr:2019kwi}. 
Contributions with one real parton in the final state were considered in refs.~\cite{Anastasiou:2013mca,Kilgore:2013gba,Duhr:2013msa,Li:2013lsa,Dulat:2014mda,Ahmed:2014pka} and the master integrals we used for our calculation were documented in refs.~\cite{Dulat:2014mda,Anastasiou:2013mca}.
Master integrals with two and three real partons were obtained for the purpose of ref.~\cite{Mistlberger:2018etf} and are based on results from refs.~\cite{Anastasiou:2014vaa,Li:2014bfa,Li:2014afw,Anastasiou:2015yha,Anastasiou:2013srw,Anastasiou:2015ema}.

We work in the $\overline{\text{MS}}$-scheme in conventional dimensional regularisation. 
The only interaction vertex that involves an axial coupling is the vertex involving the $W$ boson.
The non-diagonal flavour-structure of the charged-current interactions forces the $W$ boson to be coupled twice to same connected fermion line in interference diagrams.
As a consequence, vector and axial vector contributions to the hadronic cross section are identical and we only work with a vector current in the generation of our partonic coefficient functions.
The ultraviolet (UV) counterterm for the strong coupling constant has been determined through five loops in refs.~\cite{Tarasov:1980au,Larin:1993tp,vanRitbergen:1997va,Baikov:2016tgj,Herzog:2017ohr}.
Infrared (IR) divergences are absorbed into the definition of the PDFs using mass factorisation at N$^3$LO~\cite{Buehler:2013fha,Hoschele:2012xc,Hoeschele:2013gga}.
The mass factorisation involves convoluting lower-order partonic cross sections with the three-loop splitting functions of refs.~\cite{Moch:2004pa,Vogt:2004mw,Ablinger:2017tan}. 
We have computed all the convolutions analytically in $z$-space using the {\sc PolyLogTools} package~\cite{Duhr:2019tlz}.
After combining our interfered matrix elements with the UV and PDF-IR counterterms we send the dimensional regulator to zero and obtain our final results. Our partonic coefficients are expressed in terms of the iterated integrals in $z$ defined in ref.~\cite{Mistlberger:2018etf}. While many of these iterated integrals can be expressed in terms of harmonic or multiple polylogarithms~\cite{Remiddi:1999ew,Goncharov1998}, some integration kernels involve elliptic integrals. It is currently unknown how to express them in terms of known functions. In ref.~\cite{Mistlberger:2018etf} it was shown how to obtain fast converging series representations which allow one to achieve a relative numerical precision of (at least) $10^{-10}$ in the whole range $z\in[0,1]$ for all partonic coefficients. 

The analytic results for the partonic coefficients are provided as ancillary material with the arXiv submission. Besides the explicit analytic cancellation of the UV and IR poles, we have performed various checks to establish the correctness of our computation. First, we have checked that all logarithmic terms in the renormalisation and factorisation scales produced from the cancellation of the UV and IR poles satisfy the Dokshitzer-Gribov-Lipatov-Altarelli-Parisi (DGLAP) evolution equation~\cite{Gribov:1972ri,Altarelli:1977zs,Dokshitzer:1977sg}. Second, the limit of soft-gluon emission, which corresponds to $z\to1$, is independent of the form of the hard interaction process and only depends on the quantum numbers of the initial-state partons. Hence, the soft-virtual cross sections must be identical to the ones for neutral-current Drell-Yan production (again, apart from the contribution from the cubic Casimir operator).
We have reproduced the soft-virtual N$^3$LO cross section of refs.~\cite{Eynck:2003fn,Moch:2005ky,Ahmed:2014cla,Li:2014bfa,Catani:2014uta,H.:2020ecd}, and also the physical kernel constraints of refs.~\cite{Moch:2009hr,Soar:2009yh,deFlorian:2014vta} for the next-to-soft term of the quark-initiated cross section. 
The high-energy limit of the cross section, which corresponds to $z\to 0$, must also be identical between charged- and neutral-current production, and we have checked that our partonic cross sections have the expected behaviour in the high-energy limit~\cite{MarzaniPhD,Marzani:2008uh}.
Finally, we have checked that we reproduce the numerical results for the ratios of the inclusive $W^+$ and $W^-$ cross sections up to NNLO given in ref.~\cite{Kom:2010mv}.


\section{$W$-production at N$^3$LO}
\label{sec:pheno}

In this section we discuss our phenomenological results for inclusive (off-shell) $W$-production at N$^3$LO. 
All results were obtained in a theory with $N_f=5$ massless quark flavours (two up-type and three down-type flavours). 
The values of the mass and the width of the $W$ boson are $m_W = 80.379\,\textrm{GeV}$ and $\Gamma_W=2.085\,\textrm{GeV}$, and the vacuum expectation value is $v=246.221\,\textrm{GeV}$. The elements of the CKM matrix relevant to our computation are~\cite{Tanabashi:2018oca}:
\beq\bsp
&|V_{ud}| = 0.97446\,, \qquad |V_{us}| = 0.22452\,,\qquad |V_{ub}| = 0.00365\,,\\
&|V_{cd}| = 0.22438\,, \qquad \,|V_{cs}| = 0.97359\,,\qquad \,|V_{cb}| = 0.04214\,.
\esp\eeq
Note that only the absolute values of the entries of the CKM matrix enter our computation. 
In our phenomenological results the top quark is absent, which is equivalent to having $V_{td_i}=0$ and only considering $N_f=5$ massless degrees of freedom in loops. 
This approximation is motivated because off-diagonal CKM matrix elements are small and diagrams without a coupling of the top quark to the electroweak gauge boson decouple in the limit of infinite top quark mass. 
Corrections to this approximation, which are expected to be very small, can be computed separately and are beyond the scope of this article.
The strong coupling constant is evolved to the renormalisation scale $\mu_R$ using the four-loop QCD beta function in the $\overline{\textrm{MS}}$-scheme assuming $N_f=5$ active, massless quark flavours. Unless stated otherwise, all results are obtained for a proton-proton collider with $\sqrt{S}=13\,\textrm{TeV}$ using the zeroth member of the combined \texttt{PDF4LHC15\_nnlo\_mc} set~\cite{Butterworth:2015oua}. 

\begin{figure*}[!h]
\centering
\includegraphics[width=0.47\textwidth]{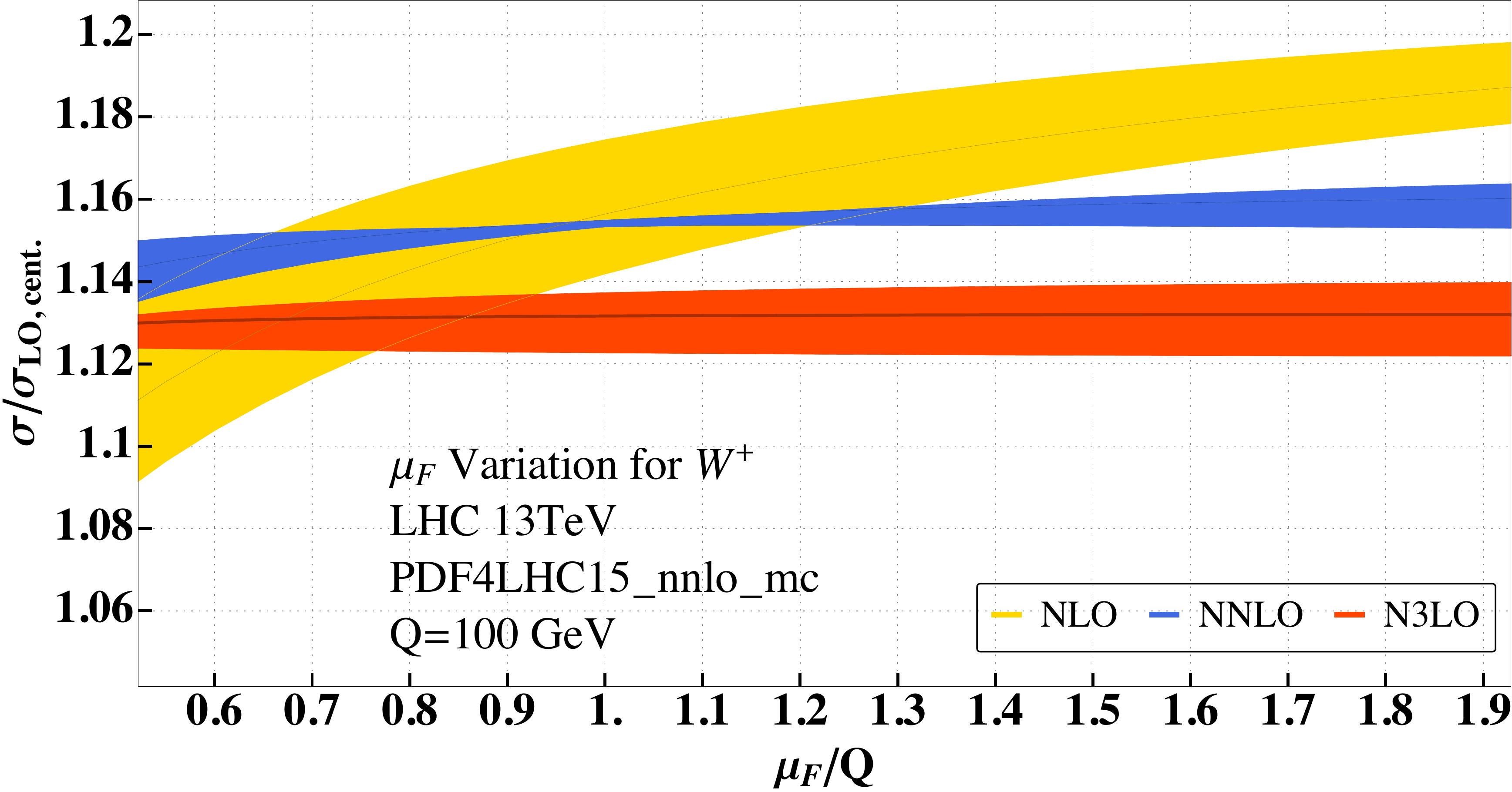}
\includegraphics[width=0.47\textwidth]{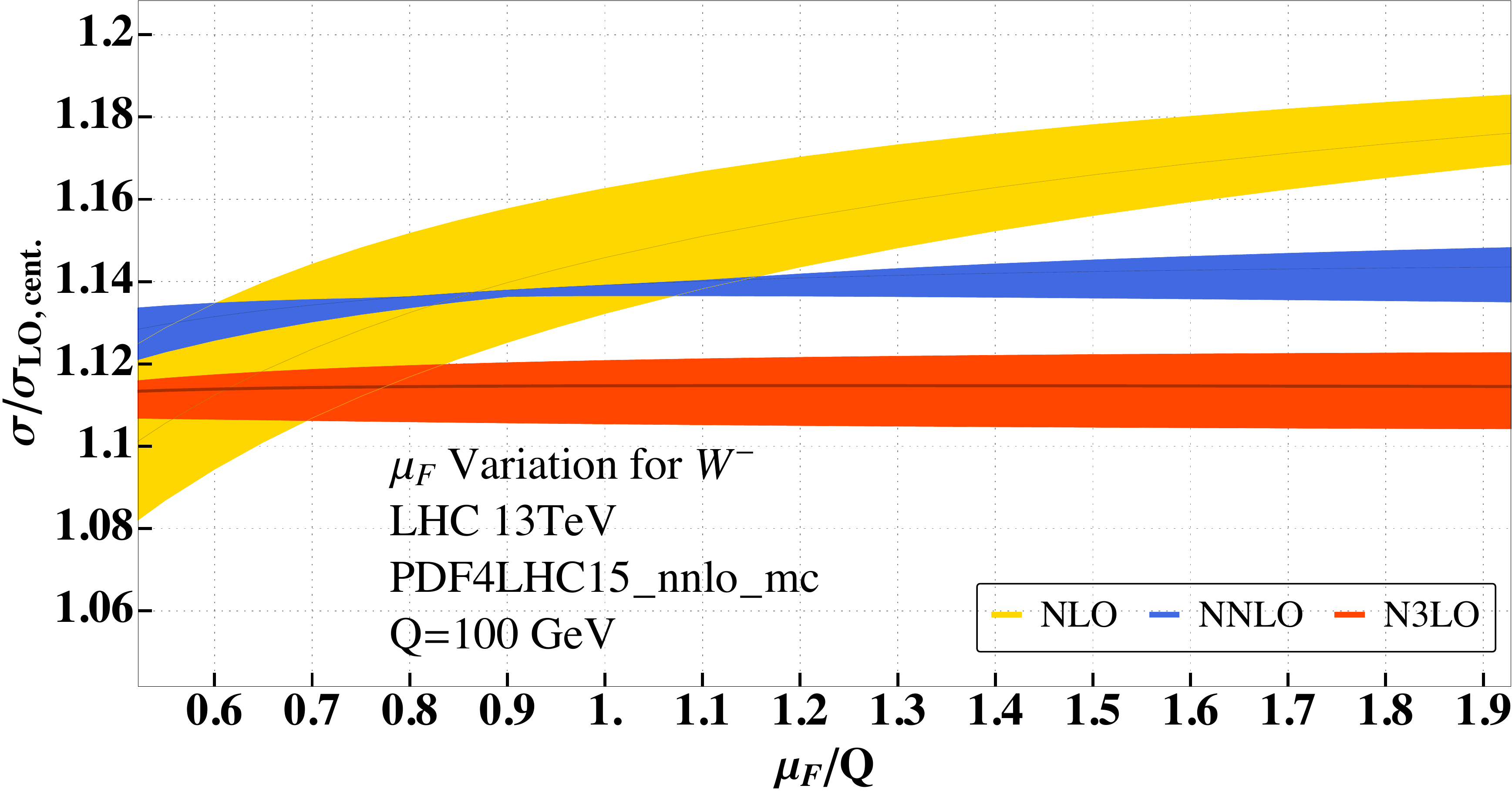}
\caption{\label{fig:muf_var}
The cross sections for producing a $W^+$ (left) or $W^-$ (right) for $\mu_R=Q=100\textrm{ GeV}$ as a function of the factorisation scale $\mu_F$. The bands are obtained by varying $\mu_R$ by a factor of 2 up and down.
The cross sections are normalised to the leading order cross section evaluated at $\mu_F=\mu_R=Q$.
}
\end{figure*}
\begin{figure*}[!h]
\centering
\includegraphics[width=0.47\textwidth]{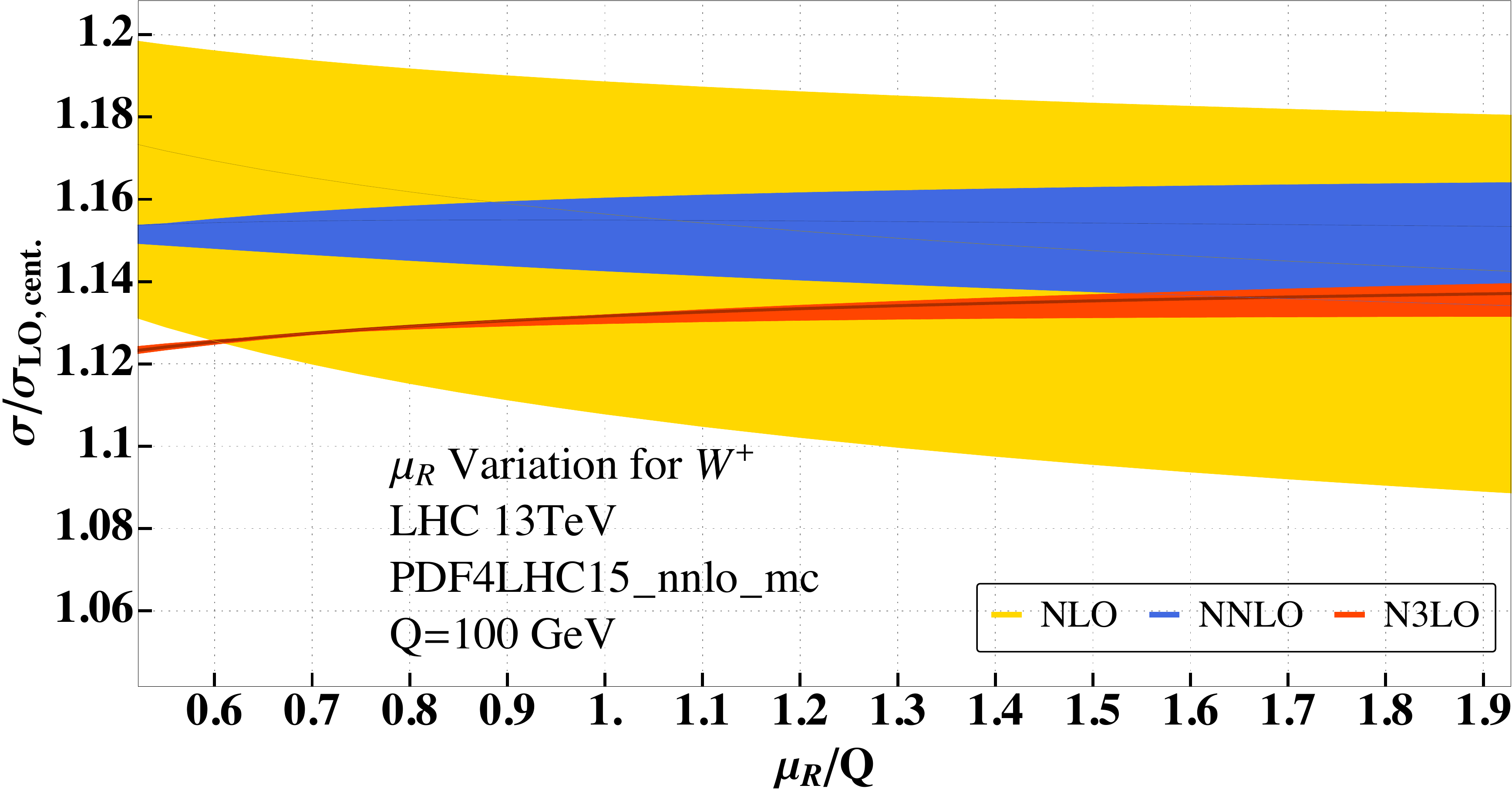}
\includegraphics[width=0.47\textwidth]{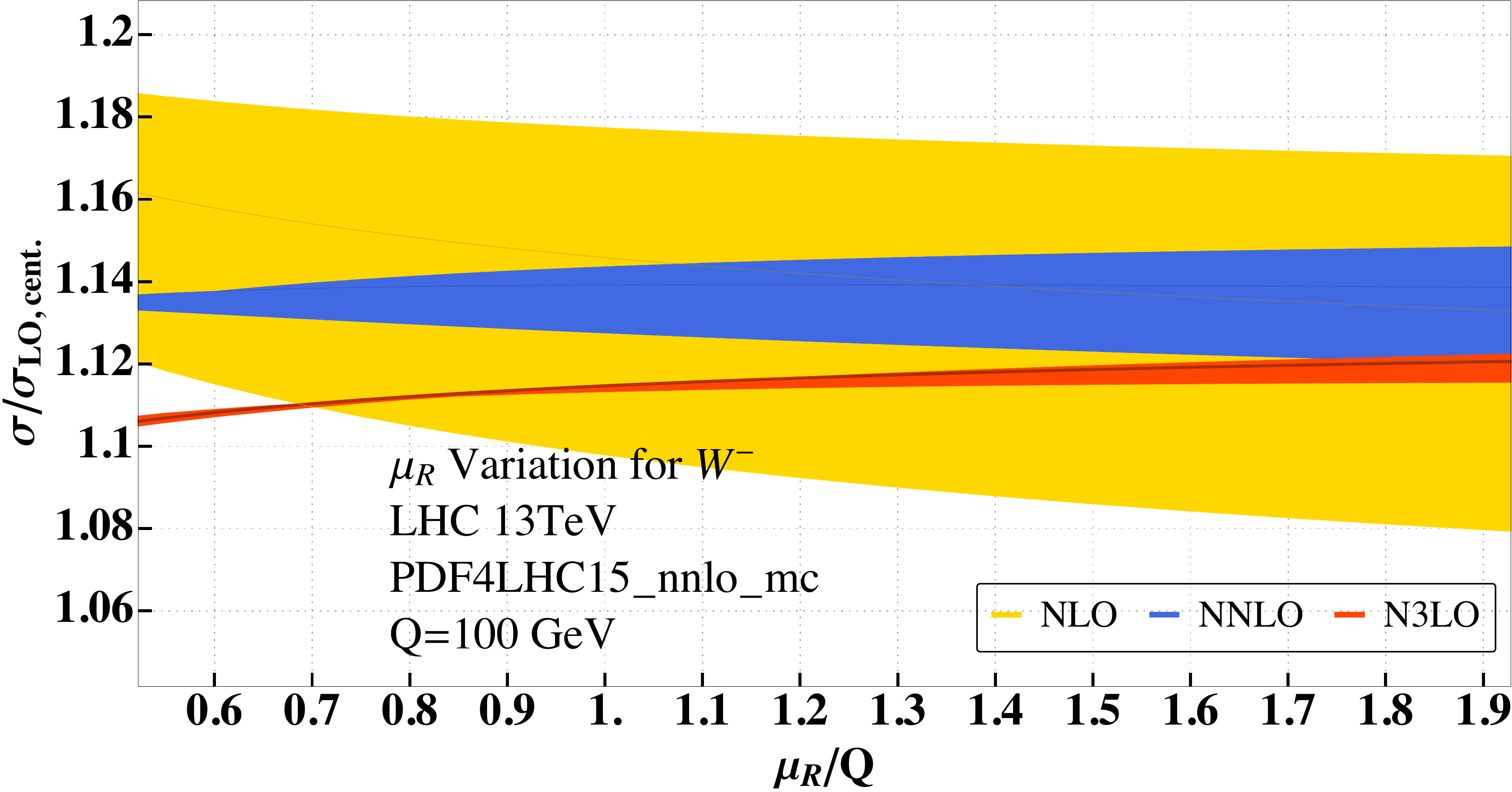}
\caption{\label{fig:mur_var}
The cross sections for producing a $W^+$ (left) or $W^-$ (right) for $\mu_F=Q=100\textrm{ GeV}$ as a function of the renormalisation scale $\mu_R$. The bands are obtained by varying $\mu_F$ by a factor of 2 up and down.
The cross sections are normalised to the leading order cross section evaluated at $\mu_F=\mu_R=Q$.
}
\end{figure*}
Figures~\ref{fig:muf_var} and~\ref{fig:mur_var} show the dependence of the fixed-order cross sections on the factorisation scale $\mu_F$ and renormalisation scale $\mu_R$, which are introduced by the truncation of the perturbative series. 
We show the variation of the cross section for $Q=100\textrm{ GeV}$ on one of the two scales with the other held fixed at $Q$. We observe that the dependence on the perturbative scales is substantially reduced as we increase the perturbative order. The dependence on the scales looks very similar to the case of the N$^3$LO cross section for the neutral-current process studied in ref.~\cite{Duhr:2020seh}.
We notice, that the dependence of the cross section on the renormalisation scale is slightly larger than on the factorisation scale.

Next, in order to quantify the size of the N$^3$LO corrections, we investigate the K-factors:
\beq\bsp\label{eq:K-factor}
\textrm{K}_{W^{\pm}}^{\text{N$^3$LO}}(Q)&\,=\frac{\sigma^{(3)}_{W^\pm}(\mu_F=\mu_R=Q)}{\sigma^{(2)}_{W^{\pm}}(\mu_F=\mu_R=Q)}\,,\\
\delta(\textrm{scale})&\,=\frac{\delta_{\textrm{scale}}(\sigma^{(3)}_{W^{\pm}})}{\sigma^{(3)}_{W^{\pm}}(\mu_F=\mu_R=Q)},
\esp\eeq
where $\sigma_{W^{\pm}}^{(n)}(\mu_F=\mu_R=Q)$ is the hadronic cross section including perturbative corrections up to $n^{\text{th}}$ order evaluated for $\mu_F=\mu_R=Q$ and $\delta_{\textrm{scale}}(\sigma^{(n)}_{W^{\pm}})$ is the absolute uncertainty on the cross section from varying $\mu_F$ and $\mu_R$ independently by a factor of two up and down around the central scale $\mu^{\textrm{cent}}=Q$ such that $\frac{1}{2}\le \frac{\mu_F}{\mu_R}\le 2$ (7-point variation).

\begin{table}[!h]
\begin{equation}
\begin{array}{c|c|c|c}\hline\hline
    Q/\textrm{GeV} &\hspace{1cm} {\textrm{K}_{W^+}^{\text{N$^3$LO}}} \hspace{1cm} & \hspace{1cm} {\textrm{K}_{W^-}^{\text{N$^3$LO}}} \hspace{1cm} & \hspace{1cm} {\textrm{K}_{\gamma^\ast}^{\text{N$^3$LO}}}  \hspace{1cm}\\
\hline
 30 & 0.953_{-1.7 \%}^{+2.5\%} & 0.950_{-1.6 \%}^{+2.6\%} & 0.952_{-1.5 \%}^{+2.5\%} \\ \hline
 50 & 0.966_{-1.2 \%}^{+1.5\%} & 0.964_{-1.2 \%}^{+1.6\%} & 0.966_{-1.1 \%}^{+1.6\%} \\ \hline
 70 & 0.974_{-1.0 \%}^{+1.1\%} & 0.972_{-0.9 \%}^{+1.2\%} & 0.973_{-0.9 \%}^{+1.1\%} \\ \hline
 80.379 & 0.976_{-0.9 \%}^{+1.0\%} & 0.975_{-0.9 \%}^{+1.0\%} & 0.976_{-0.8 \%}^{+1.0\%} \\ \hline
 90 & 0.978_{-0.8 \%}^{+0.9\%} & 0.977_{-0.8 \%}^{+0.9\%} & 0.978_{-0.7 \%}^{+0.9\%} \\ \hline
 110 & 0.981_{-0.7 \%}^{+0.7\%} & 0.980_{-0.7 \%}^{+0.8\%} & 0.981_{-0.6 \%}^{+0.7\%} \\ \hline
 130 & 0.984_{-0.6 \%}^{+0.6\%} & 0.982_{-0.6 \%}^{+0.6\%} & 0.983_{-0.6 \%}^{+0.6\%} \\ \hline
 150 & 0.985_{-0.5 \%}^{+0.5\%} & 0.984_{-0.5 \%}^{+0.6\%} & 0.985_{-0.5 \%}^{+0.5\%} \\ \hline
 500 & 0.997_{-0.1 \%}^{+0.2\%} & 0.996_{-0.1 \%}^{+0.2\%} & 0.996_{-0.1 \%}^{+0.2\%} \\ \hline
 800 & 0.999_{-0.05 \%}^{+0.09\%} & 0.999_{-0.05 \%}^{+0.1\%} & 0.999_{-0.04 \%}^{+0.09\%} \\ \hline
\hline
\end{array}
\nonumber
\end{equation}
\caption{\label{tab:K-factors} The QCD K-factor at N$^3$LO for charged-current and neutral-current Drell-Yan production. All central values and uncertainties are computed according to eq.~\eqref{eq:K-factor}. The results for neutral-current Drell-Yan production are taken from ref.~\cite{Duhr:2020seh}. }
\end{table}
Table~\ref{tab:K-factors} shows the results for the K-factors for both charged- and neutral-current Drell-Yan production for various values of $Q$ (the results for the neutral-current process are taken from ref.~\cite{Duhr:2020seh}). 
We observe that the K-factor reaches up to 5\% for small values of $Q$, and the impact of the N$^3$LO corrections gets smaller as $Q$ increases. Moreover, we find that the K-factors at N$^3$LO are identical between the charged- and neutral-current processes, confirming and extending results at lower orders, and hinting towards a universality of QCD corrections to vector-boson production in proton collisions.

\begin{figure*}[!h]
\centering
\includegraphics[width=0.47\textwidth]{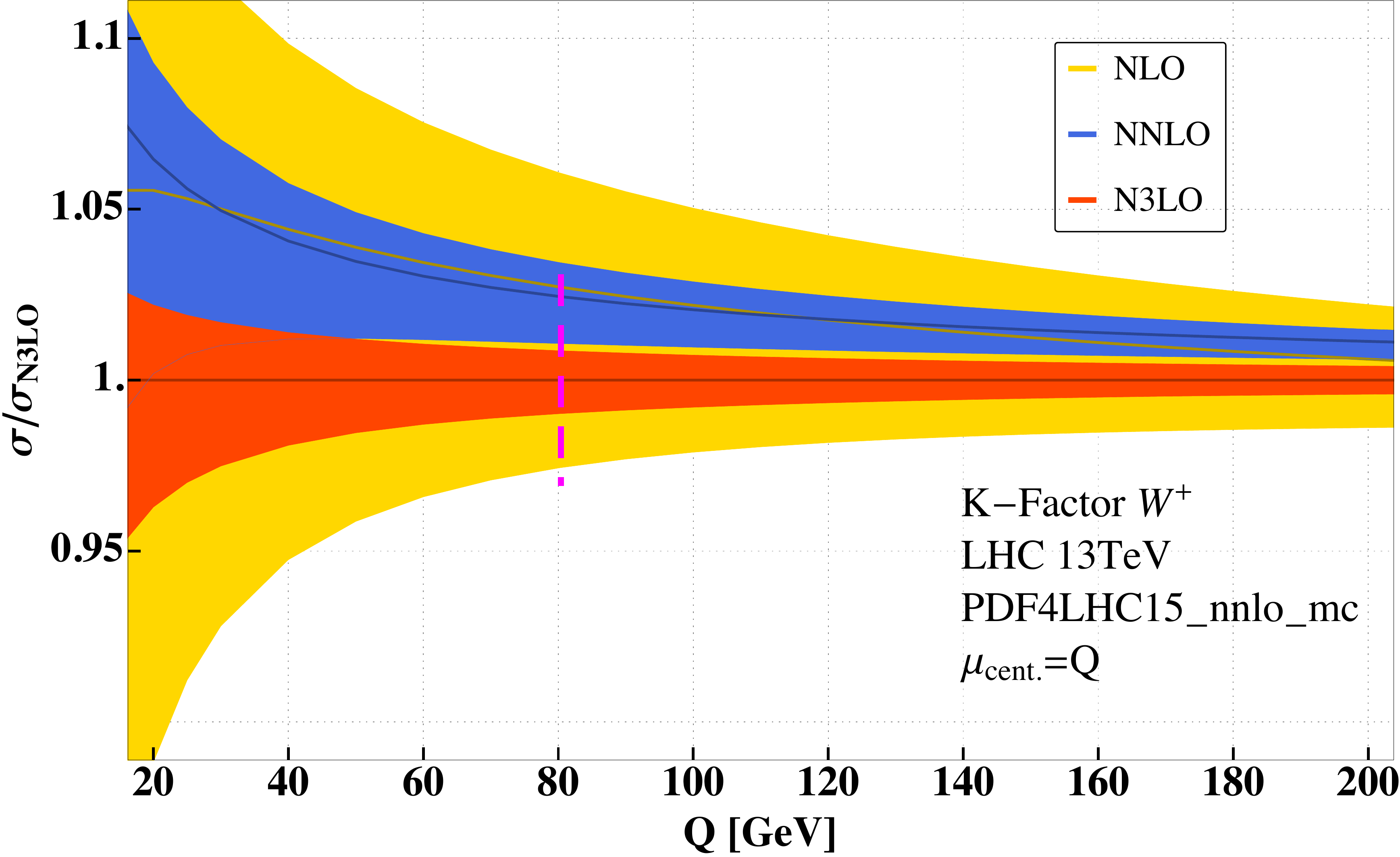}
\includegraphics[width=0.47\textwidth]{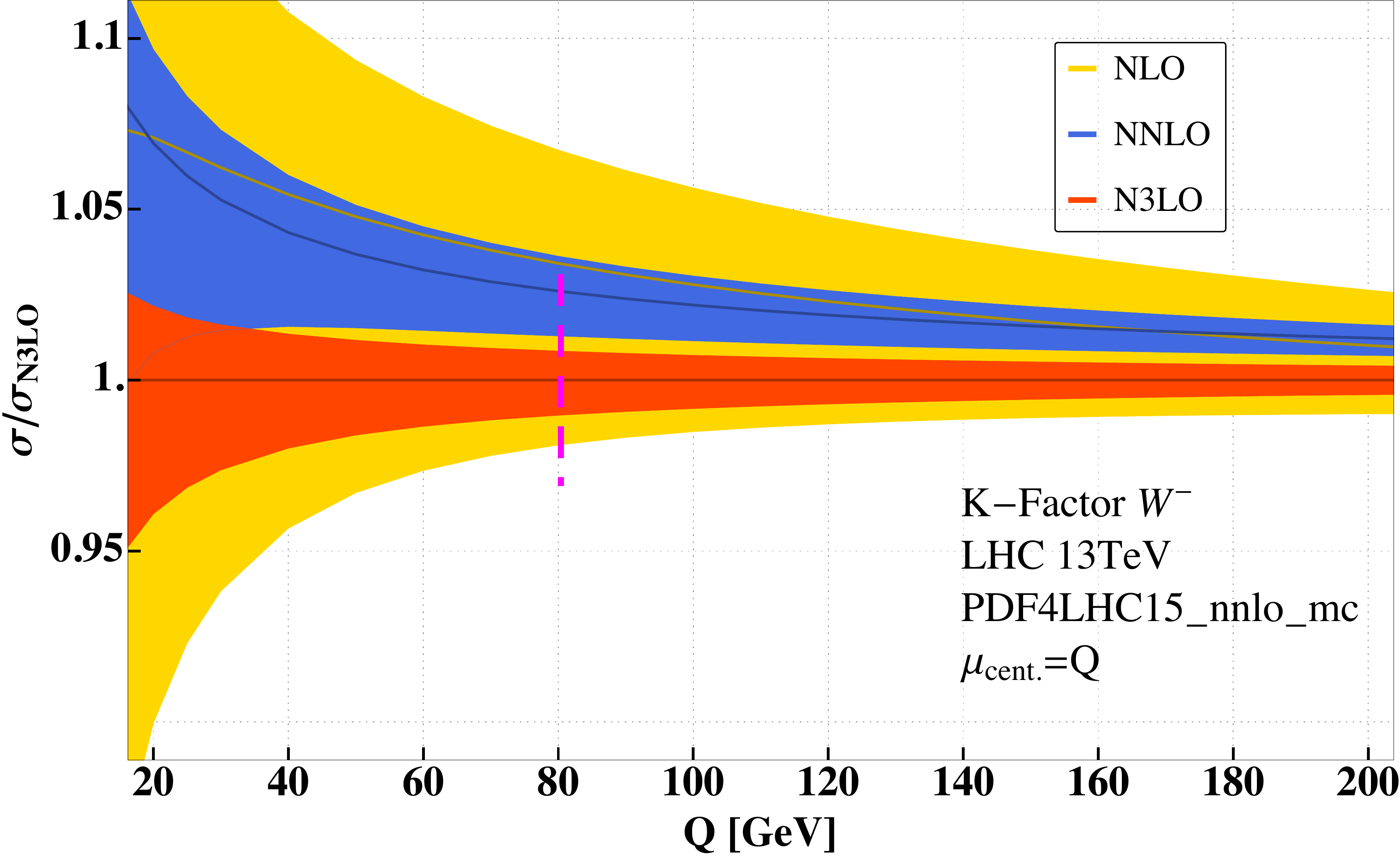}
\caption{\label{fig:Q_variation_Q}
The cross sections for producing a $W^+$ (left) or $W^-$ (right)  as a function of the virtuality $Q$ normalised to the N$^3$LO prediction. 
The uncertainty bands are obtained by varying $\mu_F$ and $\mu_R$ around the central scale $\mu^{\textrm{cent}}=Q$.
The dashed magenta line indicates the physical W boson mass, $Q=m_W$.
}
\end{figure*}
Figure~\ref{fig:Q_variation_Q} shows the values of the cross section normalised to the N$^3$LO cross section as a function of the virtuality $Q$. 
The uncertainty bands are obtained by varying the renormalisation and factorisation scales independently up and down as described above around the central scale $\mu^{\textrm{cent}}=Q$. 
We observe that for $Q\gtrsim 50\,\textrm{GeV}$ the scale-variation bands at NNLO and N$^3$LO do not overlap.
A similar feature was already observed for virtual photon production in ref.~\cite{Duhr:2020seh}, hinting once more towards a universality of the QCD corrections to these processes. 

\begin{figure*}[!h]
\centering
\includegraphics[width=0.47\textwidth]{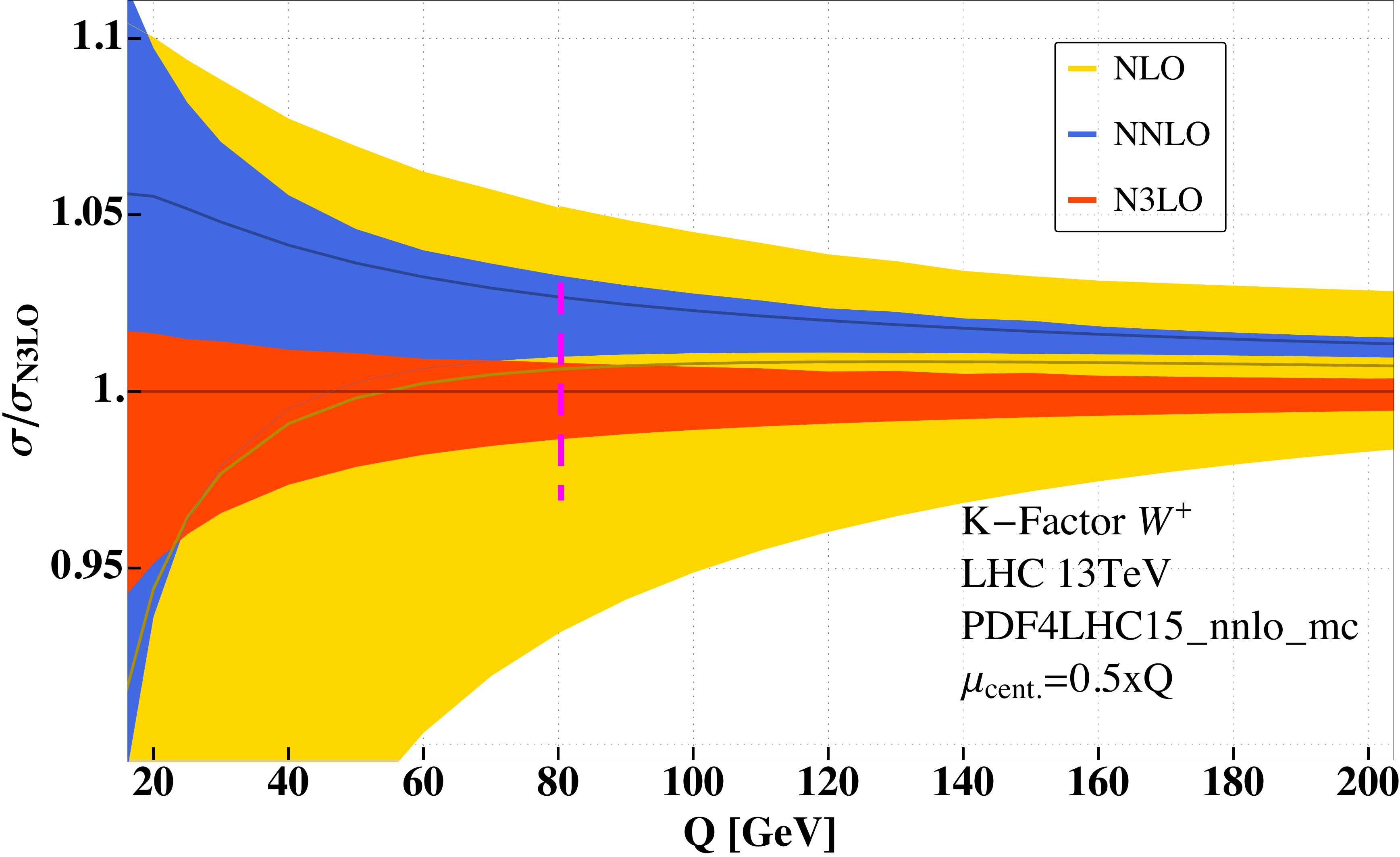}
\includegraphics[width=0.47\textwidth]{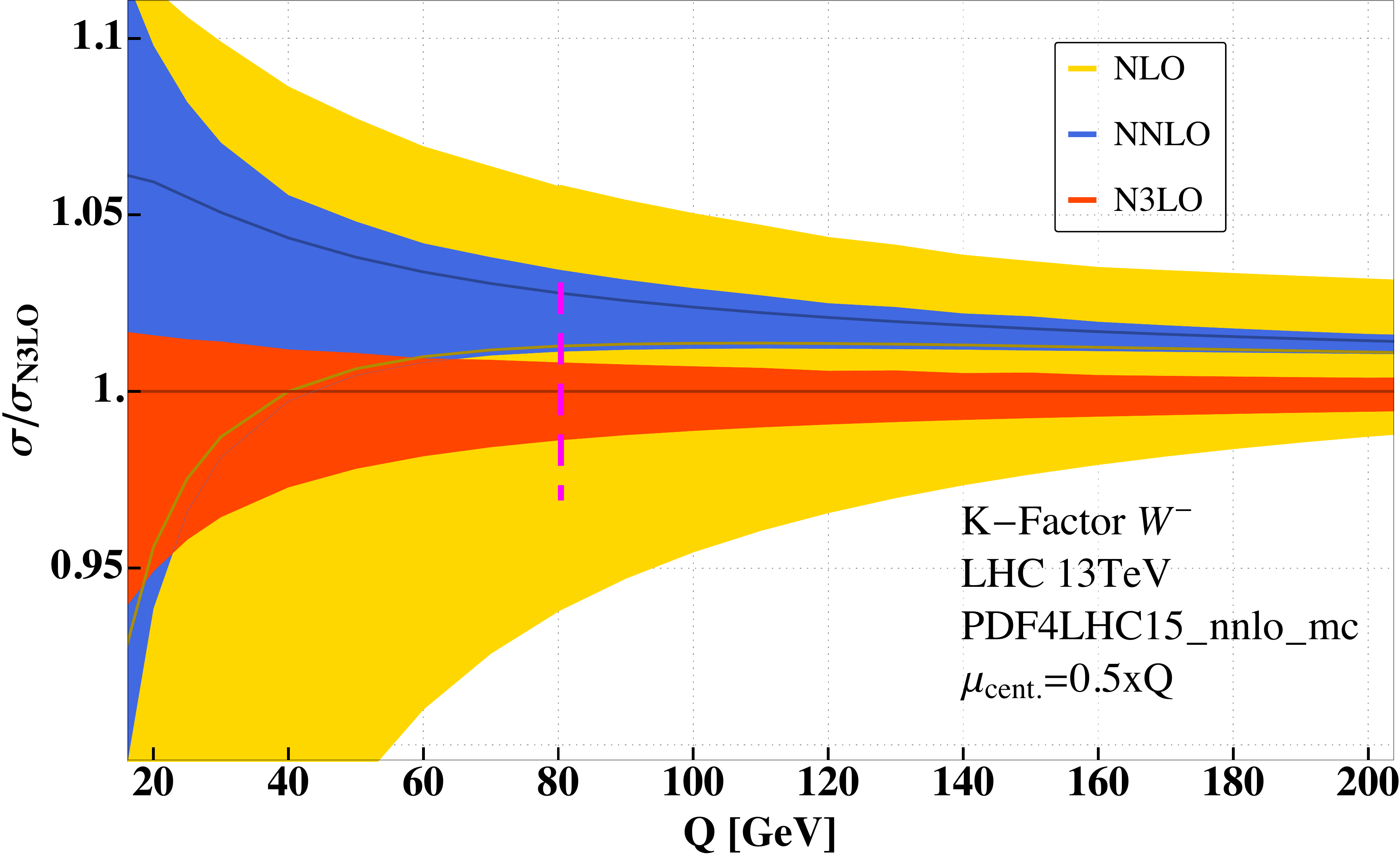}
\caption{\label{fig:Q_variation_Q/2}
The cross sections for producing a $W^+$ (left) or $W^-$ (right)  as a function of the virtuality $Q$. The uncertainty bands are obtained by varying $\mu_F$ and $\mu_R$ around the central scale $\mu^{\textrm{cent}}=Q/2$.
The dashed magenta line indicates the physical W boson mass, $Q=m_W$.
}
\end{figure*}
Figure~\ref{fig:Q_variation_Q/2} shows the scale variation of the cross section with a different choice for the central scale, $\mu^{\textrm{cent}}=Q/2$.
It is known that for Higgs production a smaller choice of the factorisation scale leads to an improved convergence pattern and the bands from scale variations are strictly contained  in one another. We observe here that the two scale choices share the same qualitative features. 

The fact that the scale variation bands do not overlap puts some doubt on whether it gives a reliable estimate of the missing higher orders in perturbation theory, or whether other approaches should be explored (cf.,~e.g.,~refs.~\cite{Cacciari:2011ze,Bonvini:2020xeo}). 
In ref.~\cite{Duhr:2020seh} it was noted that for virtual photon production there is a particularly large cancellation between different initial state configurations.
We observe here the same in the case of $W$ boson production. This cancellation may contribute to the particularly small NNLO corrections and scale variation bands, and it may be a consequence of the somewhat arbitrary split of the content of the proton into quarks and gluons. If these cancellations play a role in the observed perturbative convergence pattern, then it implies that one cannot decouple the study of the perturbative convergence from the structure of the proton encoded in the PDFs. We will return to this point below, when we discuss the effect of PDFs on our cross section predictions.

\begin{figure*}[!h]
\centering
\includegraphics[width=0.47\textwidth]{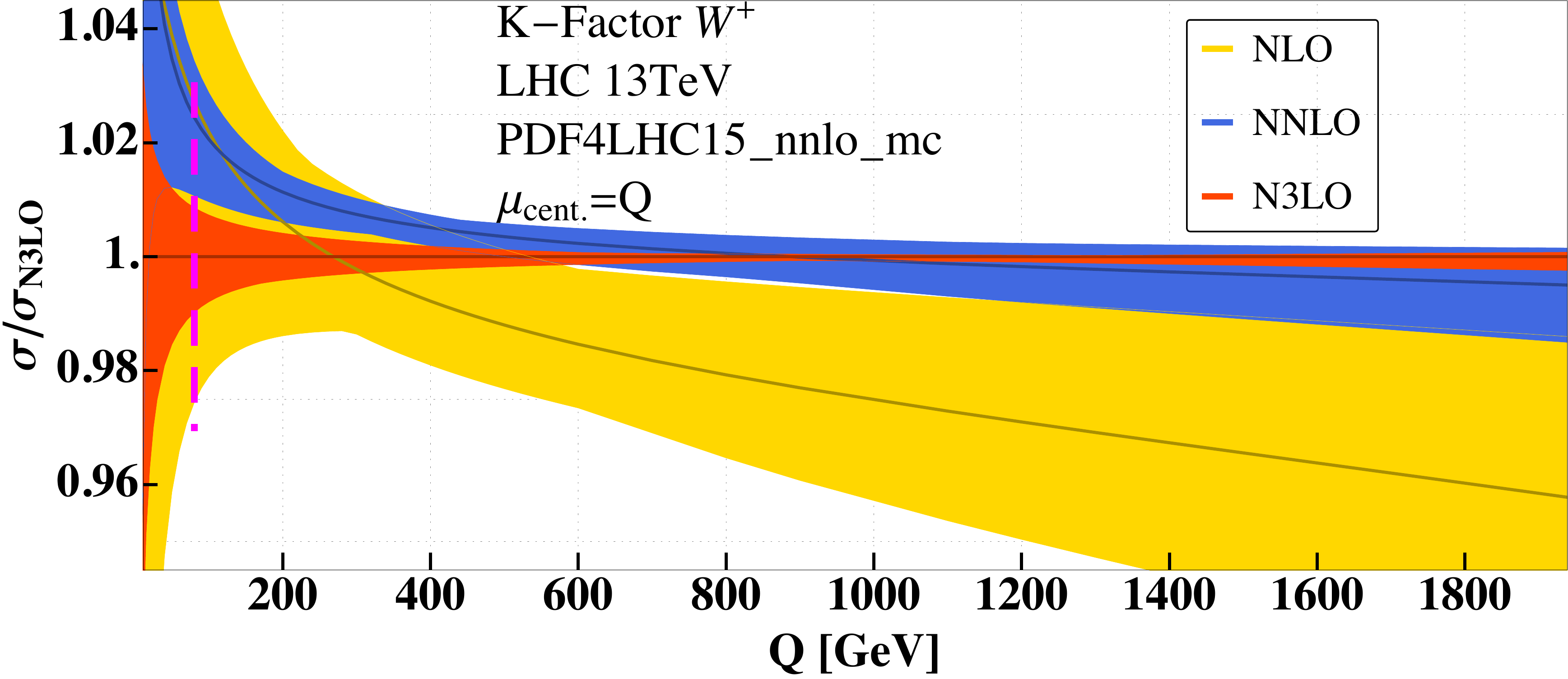}
\includegraphics[width=0.47\textwidth]{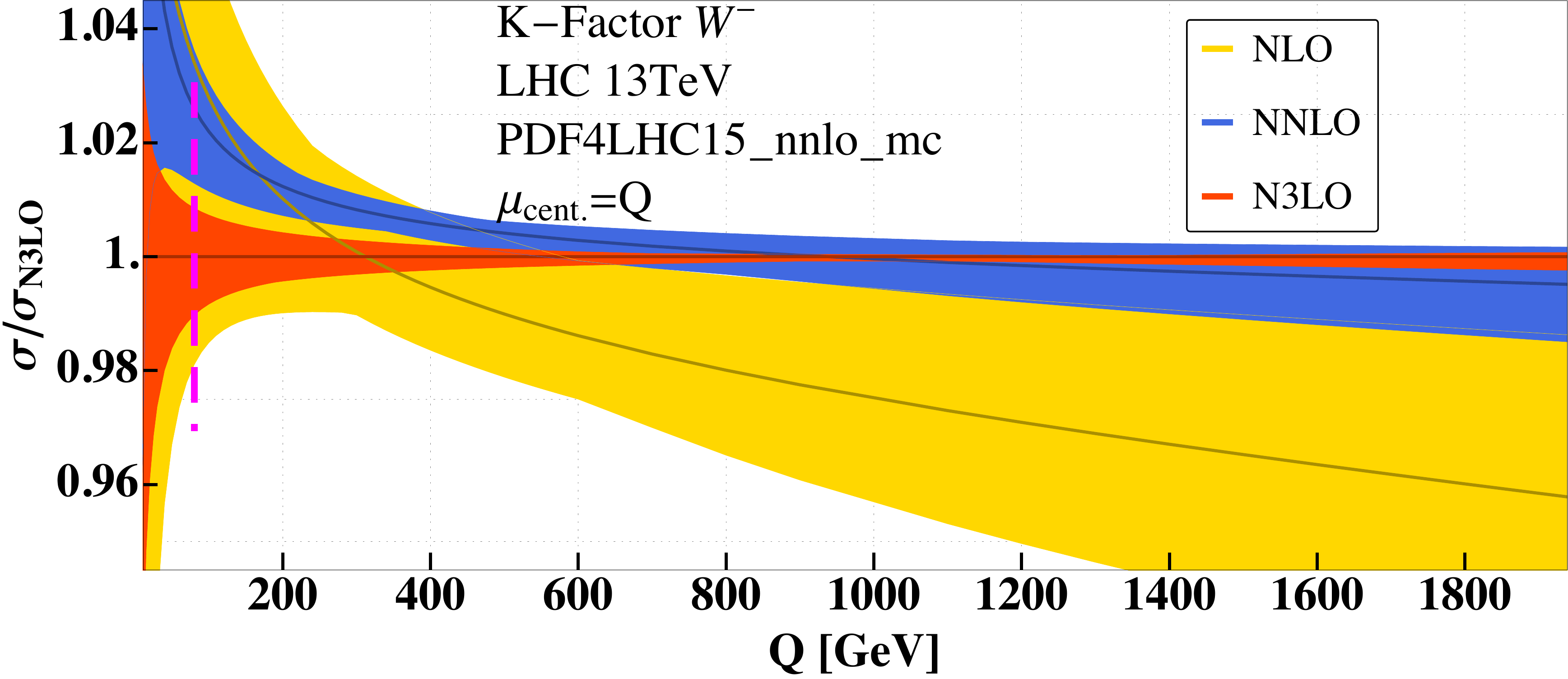}
\caption{\label{fig:QVariation_Large}
The cross sections for producing a $W^+$ (left) or $W^-$ (right)  as a function of the virtuality $Q$. The uncertainty bands are obtained by varying $\mu_F$ and $\mu_R$ around the central scale $\mu^{\textrm{cent}}=Q$.
The dashed magenta line indicates the physical W boson mass, $Q=m_W$.
}
\end{figure*}
Figure~\ref{fig:QVariation_Large} shows the production cross section for an off-shell $W$ boson normalised to the prediction at N$^3$LO for a larger range of virtualities ($Q\leq 2\,\textrm{TeV}$). 
We see that for larger values of  the virtuality ($Q> 550\,\textrm{GeV}$) the bands derived from scale variation at NNLO and N$^3$LO start to overlap.
We also observe a more typical shrinking of the scale variation bands as well as a small correction at N$^3$LO.

\begin{figure*}[!h]
\centering
\includegraphics[width=0.47\textwidth]{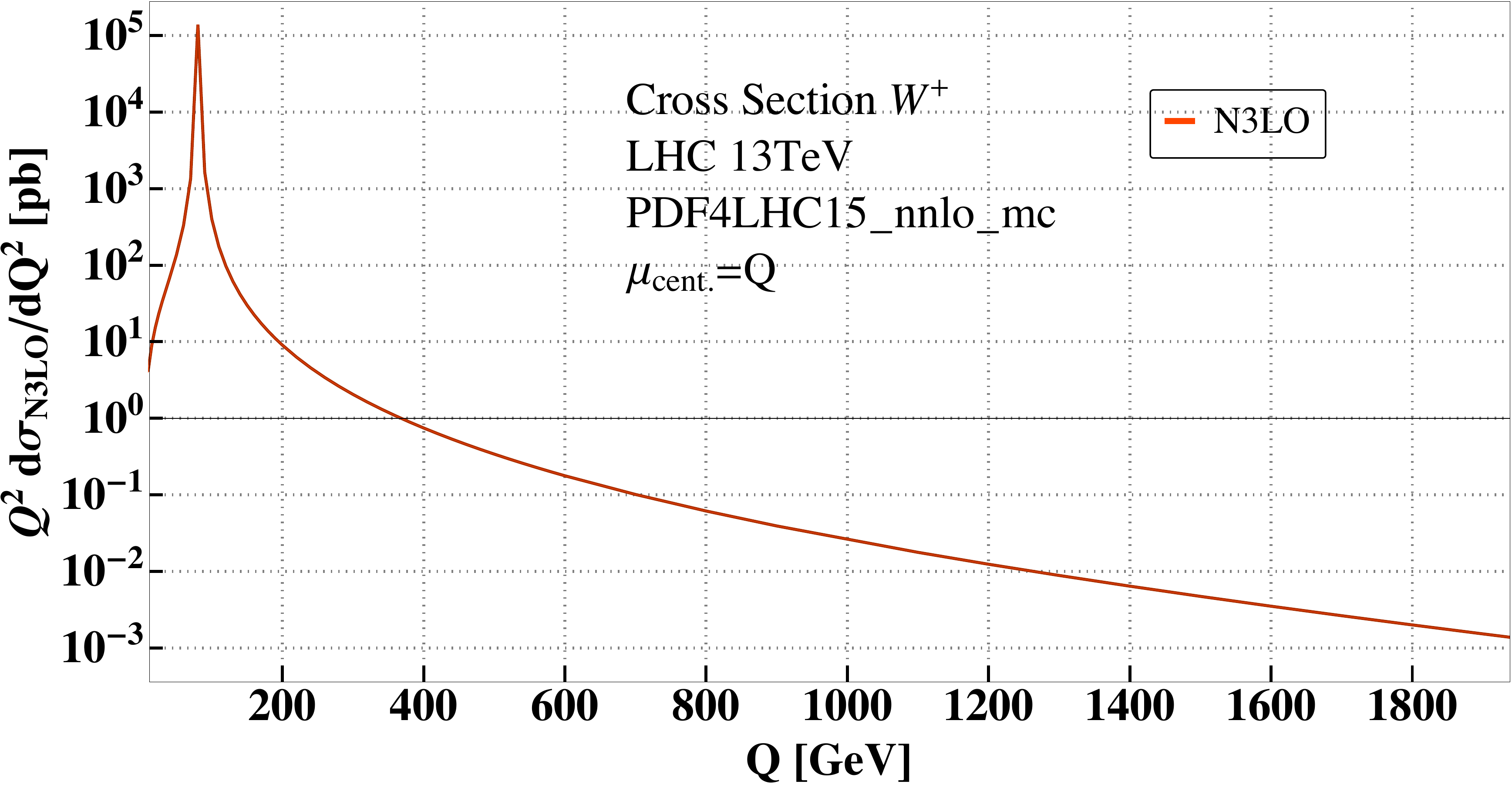}
\includegraphics[width=0.47\textwidth]{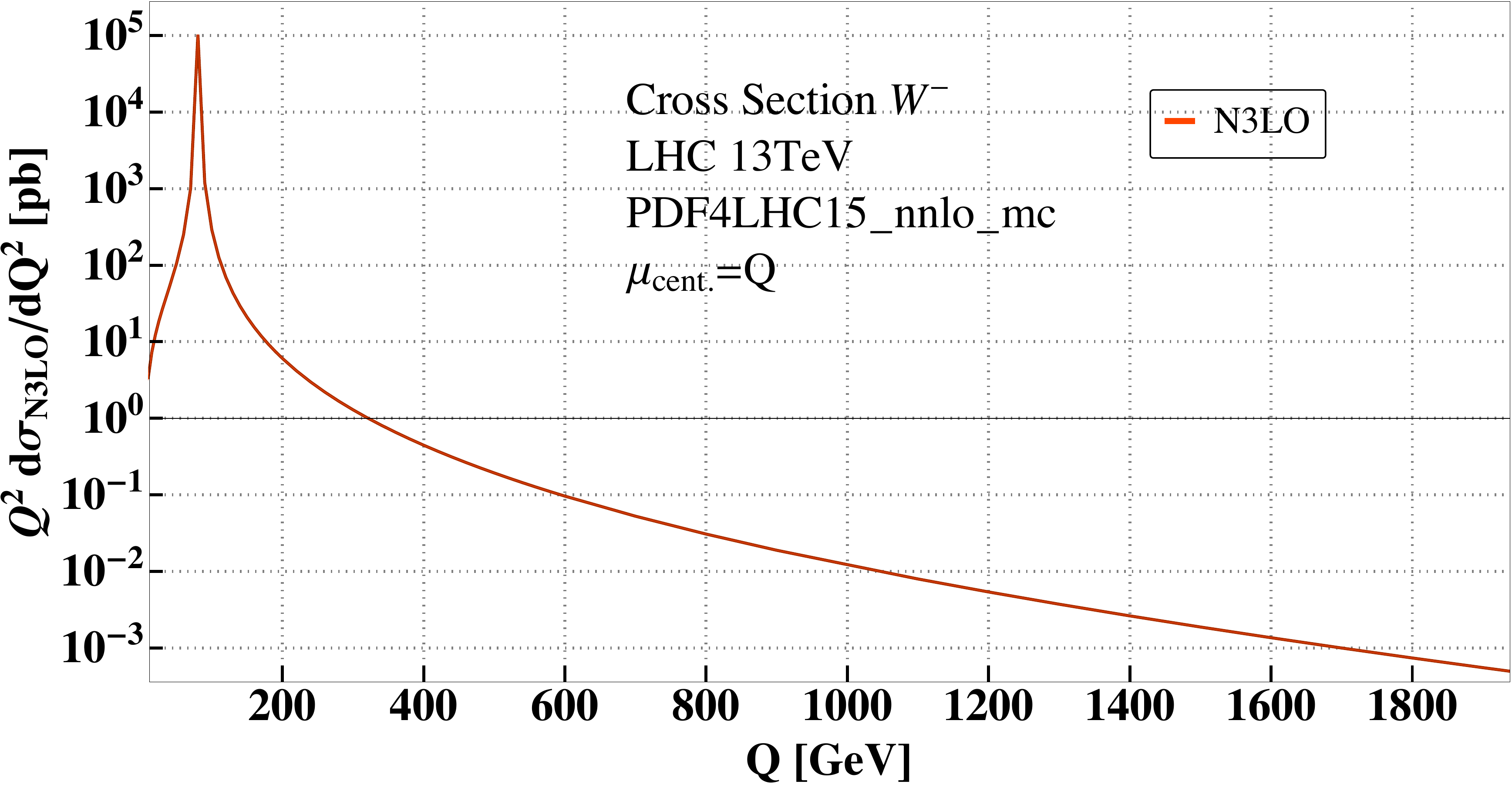}
\caption{\label{fig:NominalQVar}
The cross sections for producing a lepton-neutrino pair via an off-shell $W$ boson as a function of the invariant mass of the final state, or equivalently the virtuality of the $W$ boson, cf.~eq.~\eqref{eq:hadrdef}.
}
\end{figure*}
Figure~\ref{fig:NominalQVar} shows the nominal production cross section of a lepton-neutrino pair at the LHC at 13 TeV centre of mass energy, as defined in eq.~\eqref{eq:hadrdef}.

\begin{figure*}[!h]
\centering
\includegraphics[width=0.47\textwidth]{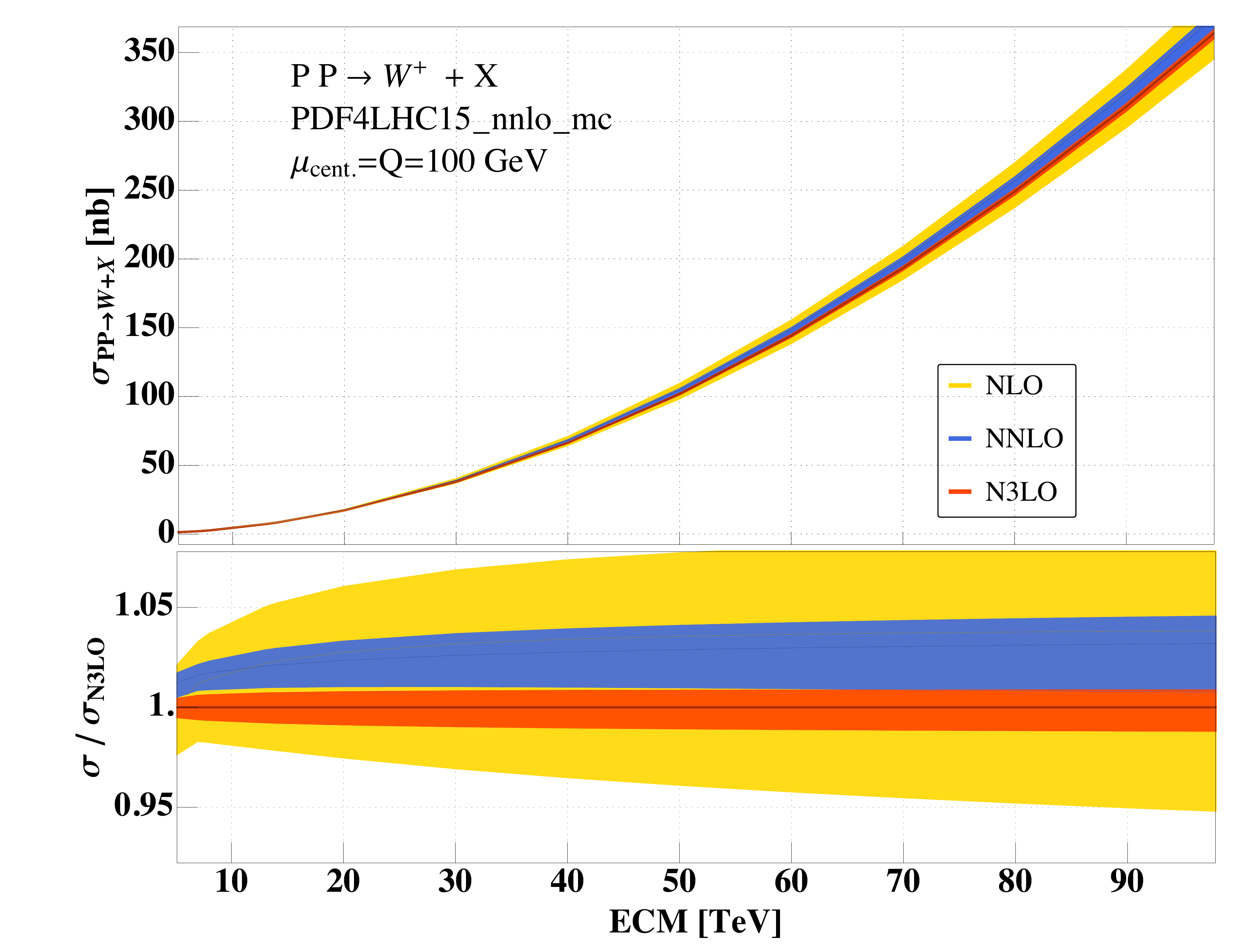}
\includegraphics[width=0.47\textwidth]{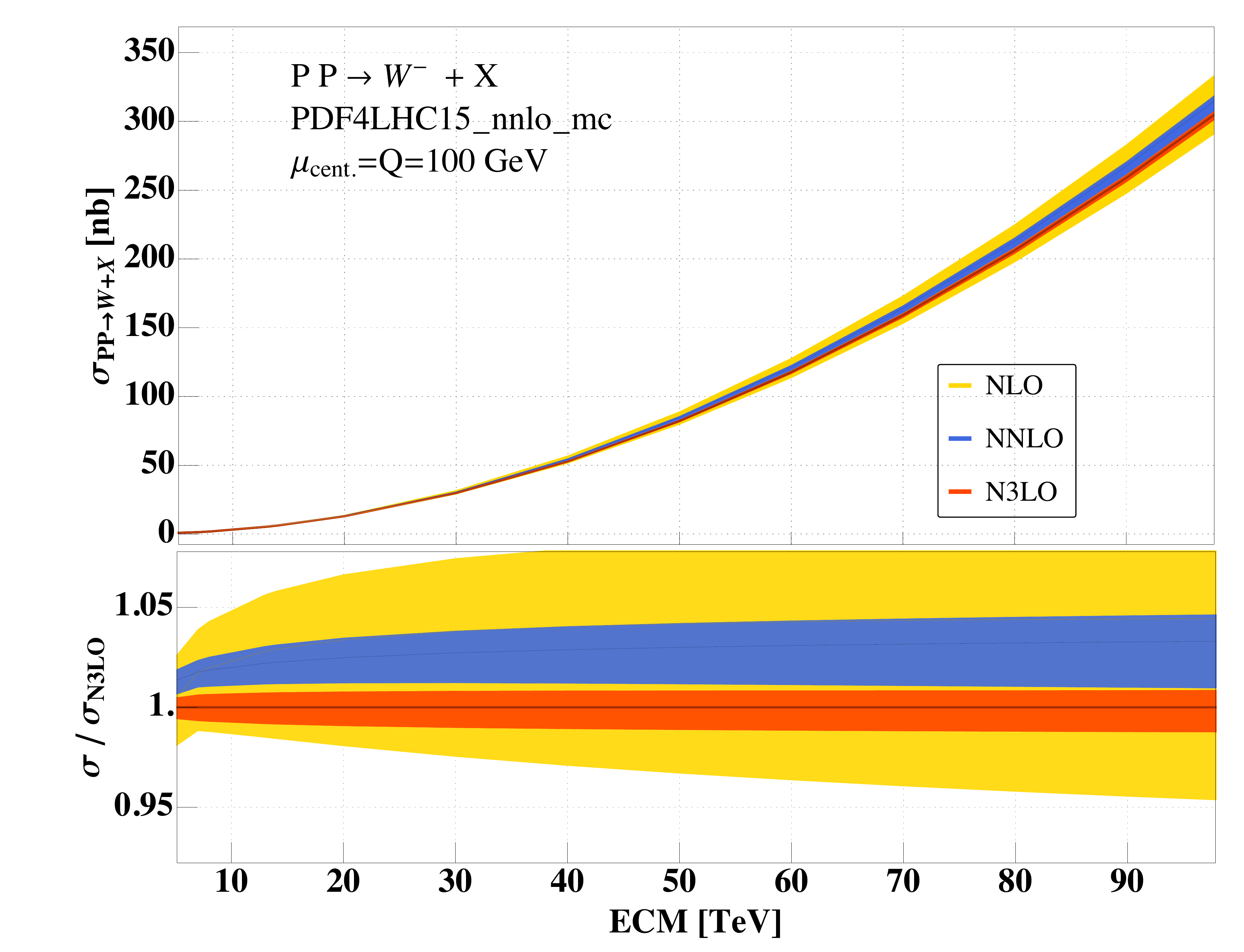}
\caption{\label{fig:ECMVariation}
The cross sections for producing a $W^+$ (left) or $W^-$ (right)  as a function of the hadronic centre of mass energy for $Q=100$ GeV.
The uncertainty bands are obtained by varying $\mu_F$ and $\mu_R$ around the central scale $\mu^{\textrm{cent}}=Q$ (see text for details).
}
\end{figure*}
Figure~\ref{fig:ECMVariation} shows the variation of K-factors as a function of the energy of the hadron collider for $Q=100$ GeV.
The orange, blue and red bands correspond to predictions with the perturbative cross section truncated at NLO, NNLO and N$^3$LO, and the size of the band is obtained by performing a 7-point variation of $(\mu_F,\mu_R)$ around the central scale $\mu^{\textrm{cent}}=Q$.
We observe that the NLO, NNLO and N$^3$LO K-factors are relatively independent of the centre of mass energy.
Furthermore, we see that the bands due to scale variation at NNLO and N$^3$LO do not overlap for a large range of center of mass energies. 
However, the gap is narrowed at the extreme end of the range of energies considered here.

Parton distribution functions are extracted from a large set of measurements and are consequently subject to an uncertainty related to the input as well as to the methodology used to extract the PDFs. 
Here, we follow the prescription of ref.~\cite{Butterworth:2015oua} for the computation of PDF uncertainties $\delta(\text{PDF})$ using the Monte Carlo method.
Furthermore, also the strong coupling constant is an input parameter for our computation.
The PDF set $\mathtt{PDF4LHC15\_nnlo\_mc}$ uses $\alpha_S=0.118$ as a central value and two additional PDF sets are available that allow for the correlated variation of the strong coupling constant in the partonic cross section and the PDF sets to $\alpha_S^{\text{up}}=0.1195$ and $\alpha_S^{\text{down}}=0.1165$. 
This sets allow us to deduce an uncertainty $\delta(\alpha_S)$ on our cross section following the prescription of ref.~\cite{Butterworth:2015oua}.
We combine the PDF and strong coupling constant uncertainties in quadrature to give
\beq
\delta(\text{PDF}+\alpha_S)=\sqrt{\delta(\text{PDF})^2+\delta(\alpha_S)^2}\,.
\eeq

In our computation we use NNLO-PDFs, because currently there is no available PDF set extracted from data with N$^3$LO accuracy. It is tantalising to speculate if the observed convergence pattern is related to the mismatch in perturbative order used for the PDFs and the partonic cross section. We estimate the potential impact of this mismatch on our cross section predictions using a prescription
introduced in ref.~\cite{Anastasiou:2016cez} that studies the variation of the NNLO cross section as NNLO- or NLO-PDFs are used. 
This defines the PDF theory uncertainty
\beq
\label{eq:PDFTH}
\delta(\text{PDF-TH})=\frac{1}{2}\left|\frac{\sigma^{\text{(2), NNLO-PDFs}}_{W^\pm}-\sigma^{\text{(2), NLO-PDFs}}_{W^{\pm}}}{\sigma^{\text{(2), NNLO-PDFs}}_{W^\pm}}\right|.
\eeq
Here, the factor $\frac{1}{2}$ is introduced as it is expected that this effect becomes smaller at N$^3$LO compared to NNLO.

\begin{figure*}[!h]
\centering
\includegraphics[width=0.47\textwidth]{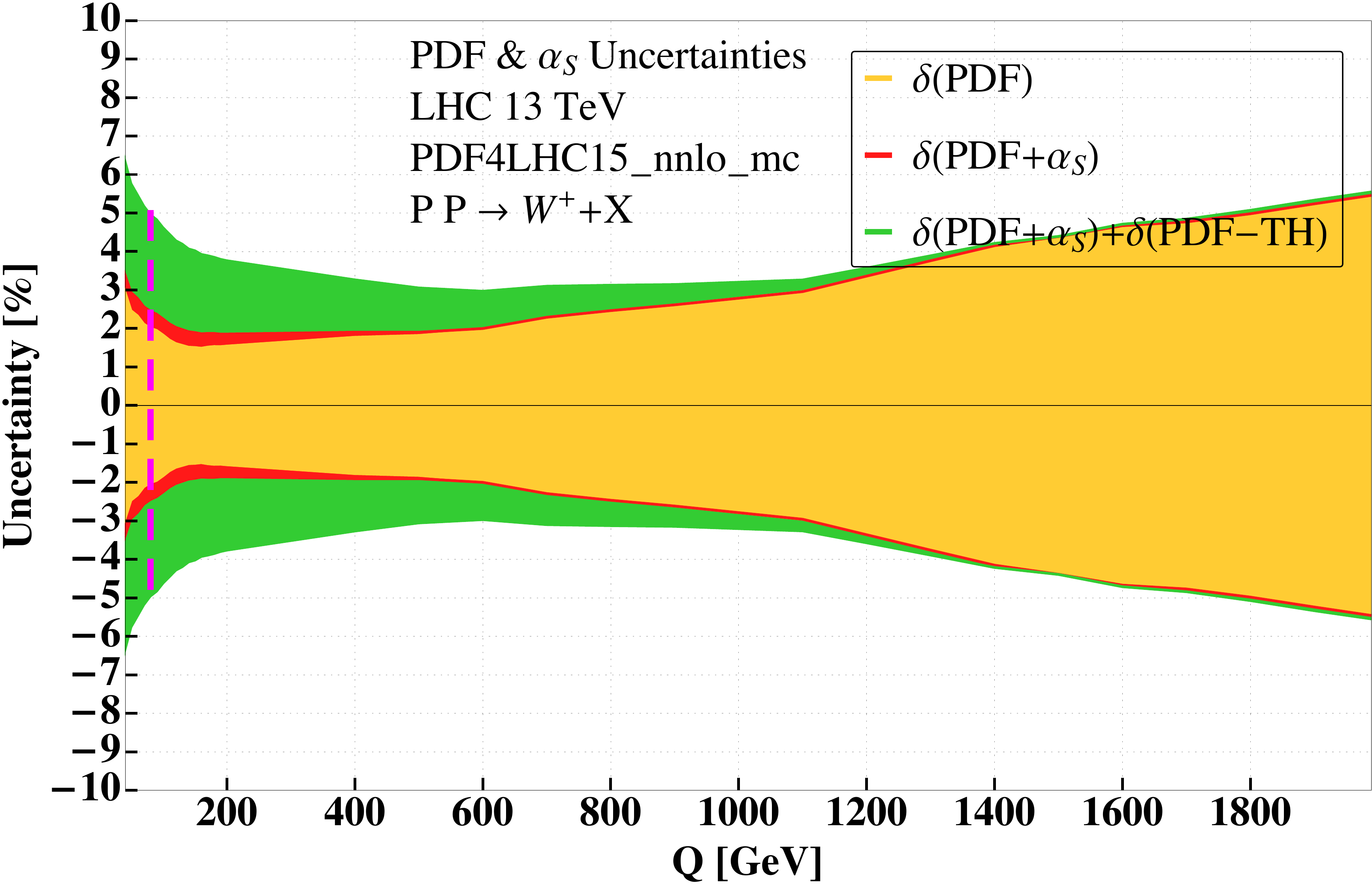}
\includegraphics[width=0.47\textwidth]{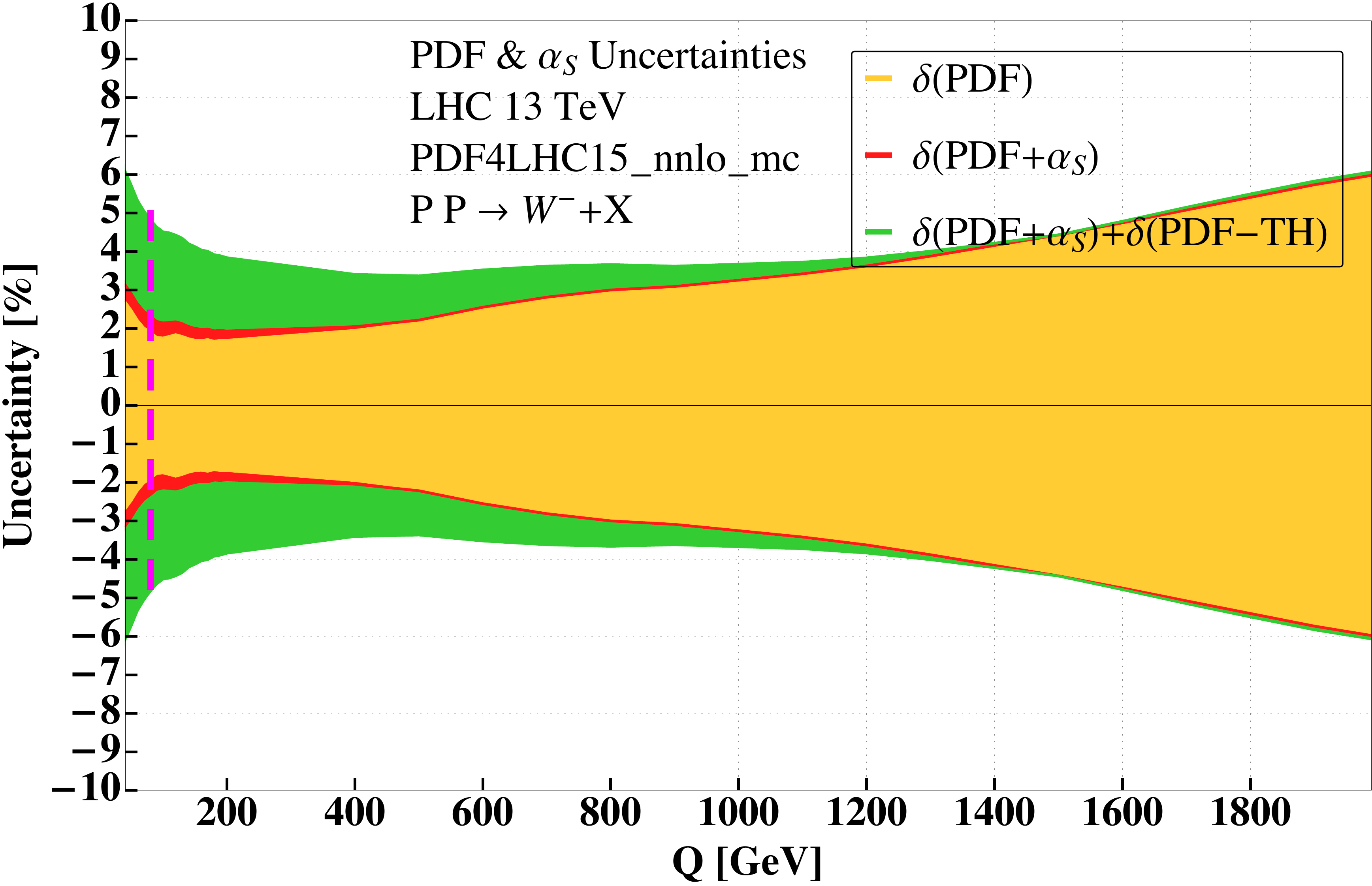}
\caption{\label{fig:Errors}
Sources of uncertainty as a function of $Q$ for the $W^+$ (left) and $W^-$ (right) K-factors. $\delta(\text{PDF})$, $\delta(\text{PDF+$\alpha_S$})$ and the sum of $\delta(\text{PDF+$\alpha_S$})$  and $\delta(\text{PDF-TH})$ are shown in orange, red and green respectively.
The dashed magenta line indicates the physical W boson mass, $Q=m_W$.
}
\end{figure*}
Figure~\ref{fig:Errors} displays the uncertainties $\delta(\text{PDF})$, $\delta(\text{PDF}+\alpha_S)$ and $\delta(\text{PDF-TH})$ as a function of $Q$ in orange, red and green respectively. 
In particular, the green band indicates the sum $\delta(\text{PDF}+\alpha_S) + \delta(\text{PDF-TH})$. 
Our findings for $\delta(\text{PDF})$ are compatible with the results of for example refs.~\cite{Ball:2017nwa,Butterworth:2015oua} where PDF effects on $W$ boson cross sections were discussed in more detail.
We observe that the estimate for $\delta(\textrm{PDF-TH})$ plays a significant role especially for low values of $Q$.
The traditional PDF uncertainty has a stronger impact for larger values of $Q$. 
Overall, we observe that the relative size of $\delta(\text{PDF})$ and $\delta(\text{PDF-TH})$ is large in comparison to the effect of varying the scales.
We conclude that future improvements in the precision of the prediction of this observable will have to tackle the problem of the uncertainties discussed here. In particular, we emphasize that
the relatively large size of $\delta(\text{PDF-TH})$ can potentially have a substantial impact on the central value of the N$^3$LO correction, especially for smaller values of $Q$.
As discussed above, there are large intricate cancellations between different initial state channels at N$^3$LO. 
This implies that a small relative change of quark vs. gluon parton densities at N$^3$LO may have an enhanced effect on the perturbative cross section as a result.
We can only wonder if the usage of true N$^3$LO parton densities could lead to N$^3$LO predictions that are fully contained in the scale variation band of the previous order. 
However, in the absence of N$^3$LO PDFs, we can only stress the importance estimating an uncertainty due to the missing N$^3$LO PDFs and suggest $\delta(\text{PDF-TH})$ as a possible estimator.


\section{Predictions for cross section ratios}
\label{sec:ratios}
In the previous section, we have seen that the conventional variation of the perturbative scales by a factor of 2 does not give a reliable estimate of the size of the missing higher orders. This motivates us to study the ratios of cross sections for the production of gauge bosons with virtuality $Q$: 
\beq\label{eq:ratio_def}
R_{XY}(Q) = \frac{\sigma_X(Q)}{\sigma_Y(Q)}\,,\qquad X,Y\in\{W^\pm,\gamma^\ast\}\,.
\eeq
Indeed, since the charged- and neutral-current Drell-Yan processes show very similar K-factors and dependences on the perturbative scales, it is conceivable that some of the uncertainties (e.g., PDF effects) cancel in the ratio, and the ratios may exhibit an enhanced perturbative stability. 
In the remainder of this section we analyse and compare different prescriptions to estimate the missing higher orders in the perturbative expansion of the cross section ratios for $\sqrt{S}= 13\, \textrm{TeV}$. We focus here on the following prescriptions:
\begin{itemize}
\item {\bf Prescription A:} We take the ratio of the perturbative expansion of the numerator and the denominator computed at a given order in perturbation theory. We choose the renormalisation and factorisation scales in the numerator and denominator in a correlated way, i.e., we always choose the same values for the scales in both the numerator and the denominator.
\item {\bf Prescription B:} Similar to Prescription A, but we do not correlate the renormalisation and factorisation scales between the numerator and the denominator. In other words, we perform independently a 7-point variation of the scales in the numerator and the denominator, and we take the envelope of the values obtained.
\item {\bf Prescription A$'$}: We choose the scales in a correlated way, but we expand the ratio in perturbation theory and only retain terms through a given order in the strong coupling.
\item {\bf Prescription B$'$:} Similar to Prescription A$'$, but we choose the scales in an uncorrelated way.
\item {\bf Prescription C:} We take the relative size of the last considered order compared to the previous one as an estimator of the perturbative uncertainty:
\beq
\delta(\text{pert.})=\pm\left|1-\frac{R_{XY}^{(n)}(Q)}{R_{XY}^{(n-1)}(Q)}\right|\times100 \%\,.
\eeq
The superscript $n$ on $R_{XY}^{(n)}(Q)$ indicates the order at which we truncate the perturbative expansion. The values of $\mu_F=\mu_R=\mu^{\textrm{cent}}$ are chosen in a correlated way in the numerator and the denominator of $R_{XY}$.
This estimator is based on the expectation that in a well-behaved perturbative expansion the subleading terms should be smaller than the last known correction. By construction, it leads to an estimate of the missing higher orders that is symmetric around the central value.
\end{itemize}
Note that Prescriptions A, A$'$, B and B$'$ are fully equivalent to all orders in perturbation theory. The truncation of the perturbative series can, however, introduce differences in the results obtained from these four prescriptions, especially at low orders in perturbation theory. 
For example, the bands obtained by varying the scales will always be larger if the scales are varied  in an uncorrelated way, because in that case the size of the band is obtained by taking the envelope of a strictly larger set of values.

\begin{table}[!h]
\begin{equation}
\begin{array}{c|cc|cc|cc}\hline\hline
   & \multicolumn{2}{c}{\textrm{NLO}} &\multicolumn{2}{c}{\textrm{NNLO}}& \multicolumn{2}{c}{\textrm{N}^3\textrm{LO}}  \\
\hline
\mu^{\textrm{cent}}& m_W & m_W/2 & m_W & m_W/2 & m_W & m_W/2\\
\hline
  \textrm{A}\phantom{'} & 1.342_{-0.08\%}^{+0.10\%}& 1.342_{-0.05\%}^{+0.07\%} & 1.348_{-0.10\%}^{+0.12\%} &1.349_{-0.11\%}^{+0.15\%} &1.350_{-0.06\%}^{+0.05\%}& 1.350_{-0.05\%}^{+0.04\%} \\
  \textrm{A}' & 1.343_{-0.16\%}^{+0.13\%}& 1.344_{-0.21\%}^{+0.10\%} & 1.349_{-0.09\%}^{+0.13\%} & 1.351_{-0.13\%}^{+0.33\%} & 1.350_{-0.03\%}^{+0.02\%} & 1.350_{-0.09\%}^{+0.01\%}\\
  \textrm{B}\phantom{'} & 1.342_{-8.08\%}^{+8.82\%} & 1.342_{-11.4\%}^{+12.9\%}&1.348_{-2.31\%}^{+2.26\%} & 1.349_{-2.27\%}^{+2.24\%}& 1.350_{-2.14\%}^{+2.21\%} &1.350_{-2.14\%}^{+2.21\%}  \\
    \textrm{B}' &1.343_{-7.40\%}^{+5.28\%} & 1.344_{-8.97\%}^{+8.09\%} & 1.349_{-2.63\%}^{+1.85\%} & 1.351_{-2.24\%}^{+2.21\%} & 1.350_{-2.25\%}^{+2.60\%} & 1.350_{-2.70\%}^{+4.65\%}\\  
    \textrm{C} &1.342_{-0.99\%}^{+0.99\%} & 1.342^{+0.58 \%}_{-0.58 \%}  & 1.349_{-0.52\%}^{+0.52\%} & 1.349^{+0.53 \%}_{-0.53 \%}  & 1.350_{-0.15\%}^{+0.15\%} &1.350^{+0.11 \%}_{-0.11 \%}  \\\hline  
    \hline
\end{array}
\nonumber
\end{equation}
\caption{\label{tab:ratio_mW} The ratio $R_{W^+W^-}$ for $Q=m_W$ computed for different values of $\mu^{\textrm{cent}}$ and with the different prescriptions mentioned in the text. }
\end{table}

Table~\ref{tab:ratio_mW} shows the prediction for $R_{W^+W^-}$  for $Q=m_W$ computed at the first few orders in perturbation theory.
First, we see that the central value is extremely stable in perturbation theory, changing only at the permille level as we go from NNLO to N$^3$LO.
The central value is pretty much independent of the choice of the central scale $\mu^{\textrm{cent}}$ and whether the ratio is expanded in perturbation theory or not (primed vs. unprimed prescriptions). 
While in general the different prescriptions lead to vastly different estimates of the missing higher orders, the predictions are similar between the primed vs. unprimed prescriptions, especially as we increase the perturbative order. This is to be expected: If the perturbative order is increased, the differences stemming from expanding or not the denominator of the ratio should decrease, which is indeed what we observe. We therefore only discuss the unprimed prescriptions from now on.
Second, we observe that Prescription B leads to an estimate that is more than an order of magnitude larger than for Prescriptions A and C. In particular, this makes one wonder if correlated scales (Prescription A \& C) tend to underestimate the size of the missing higher-order terms beyond N$^3$LO. We believe that results obtained from uncorrelated scales (Prescription B) lead to estimates that are too conservative. Indeed, since the central value of the ratio only receives permille-level corrections from NNLO to N$^3$LO and exhibits extremely good perturbative stability, one expects higher-order corrections to be at the sub-permille level, which is indeed the size of the band obtained by varying the scales in a correlated way (Prescription A \& C). It would be unreasonable to expect that the missing higher-order corrections shift the central value by 1\% or even more, which is the size of the bands obtained from the uncorrelated prescription (Prescription B). A correlated prescription is also motivated by the fact that the neutral- and charged-current processes are expected to receive very similar QCD corrections, a fact which is corroborated by the results from the previous section. Finally, we observe that Prescription C leads to an estimate that is always slightly larger (by a factor $\sim3$ for $Q=m_W$) than the one obtained from Prescription A at N$^3$LO.
We have observed that the size of the higher-order terms estimated from Prescription A always encompasses the next order in perturbation theory. 
In order words, Prescription A appears to account for the effect of missing higher orders even though it estimates relatively small residual uncertainties for the point $Q=m_W$. 
Below we study ratios of cross sections as a function of a range of different values of $Q$ allowing us to comment on Prescription A in more detail.

\begin{figure*}[!h]
\centering
\includegraphics[width=0.32\textwidth]{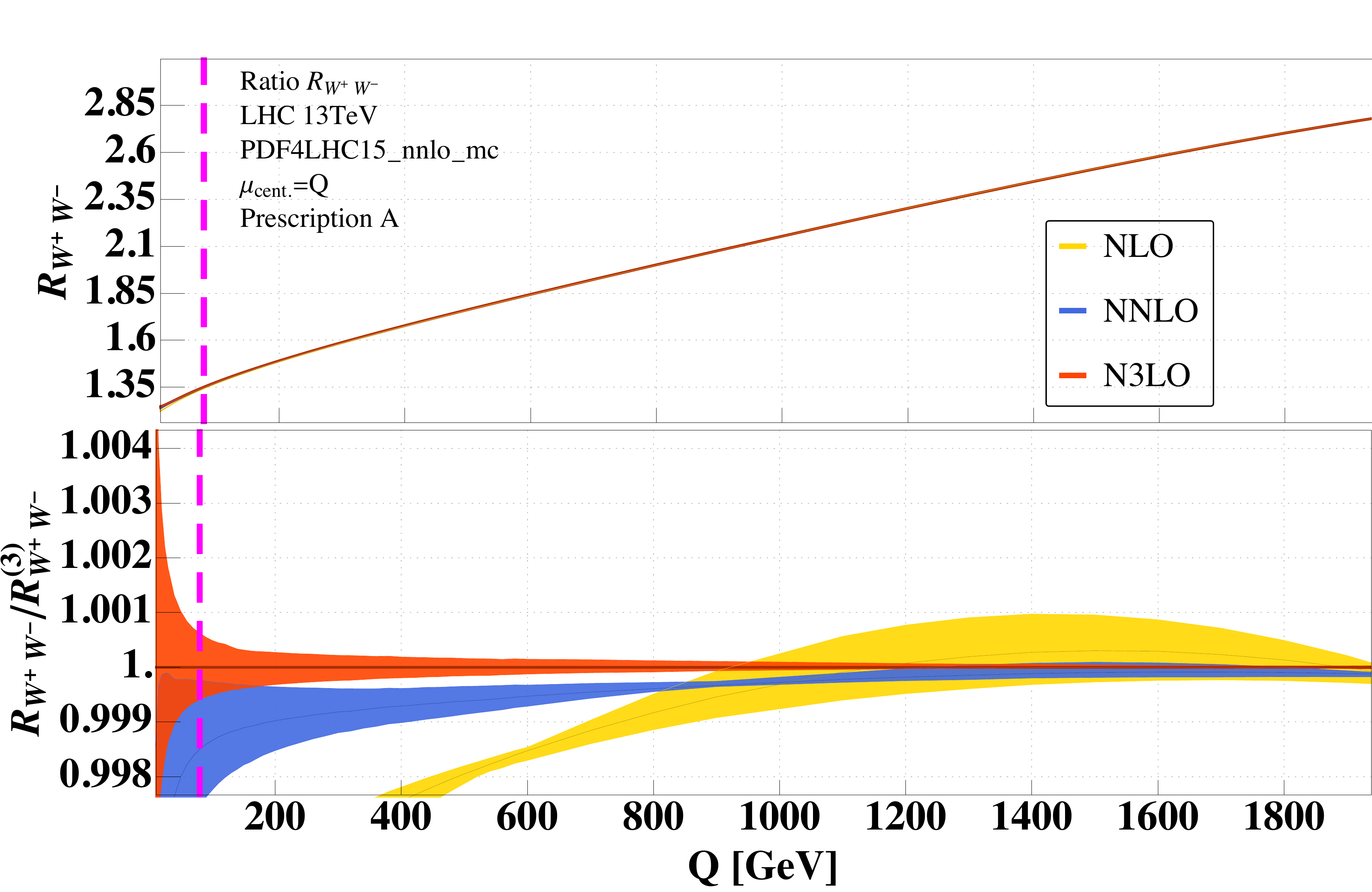}\includegraphics[width=0.32\textwidth]{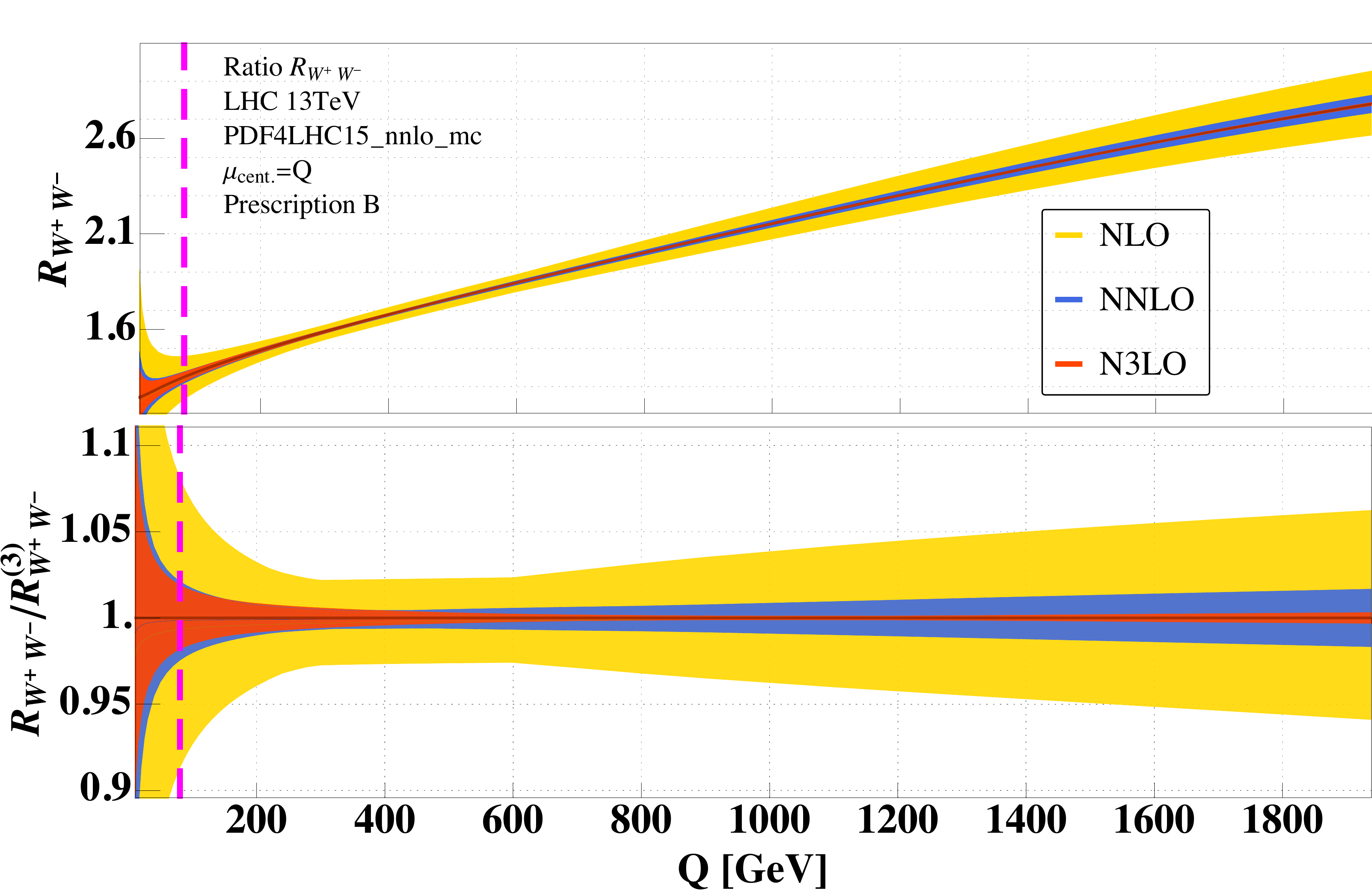}
\includegraphics[width=0.32\textwidth]{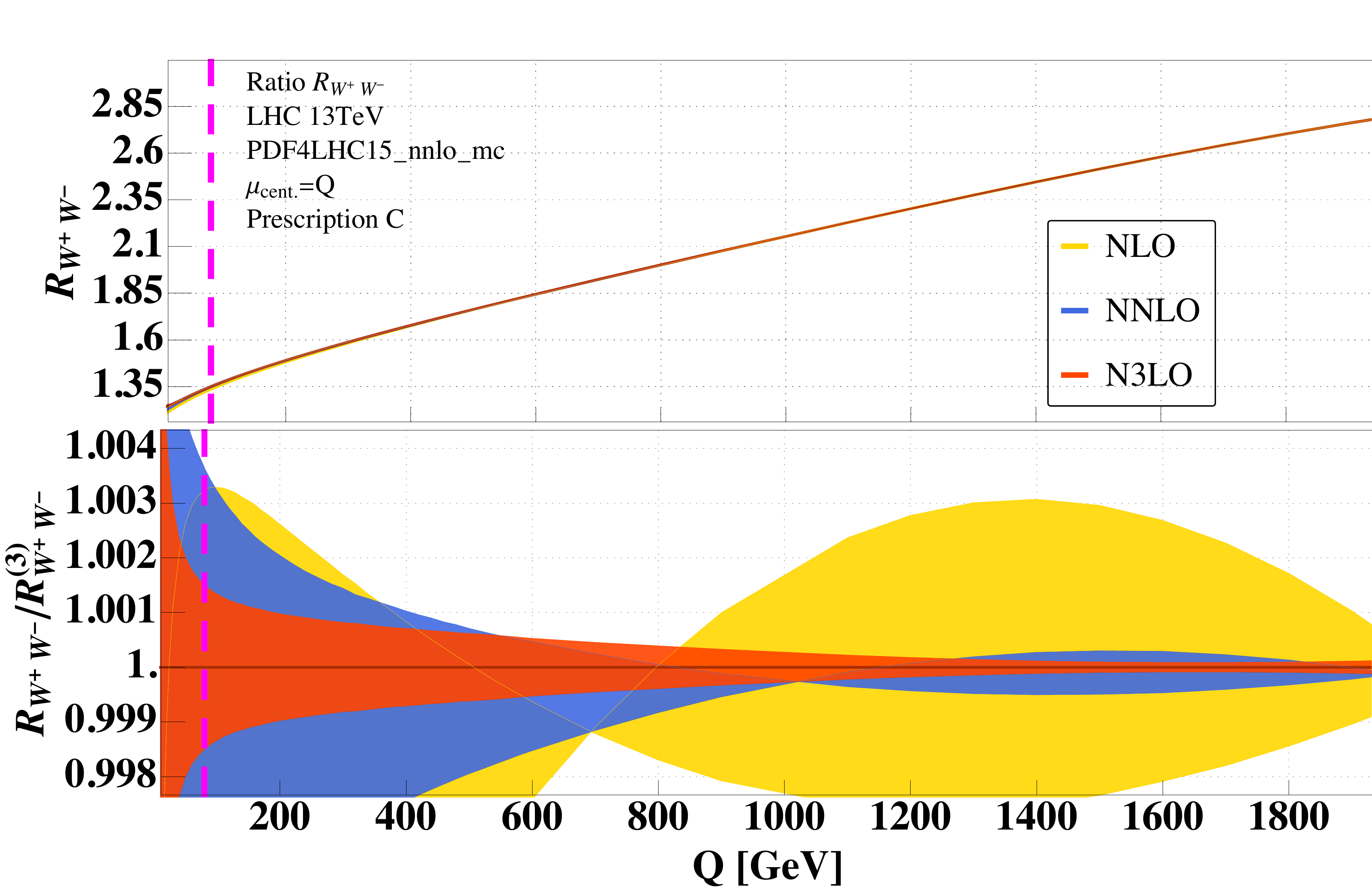}
\caption{\label{fig:PrescriptionComparison}
The upper panels show the ratio $R_{W^+W^-}$ (left) with bands computed with Prescription A (left), Prescription B (middle) and Prescription C (right).
The lower panels show the same normalised to the value of the ratio computed at N$^3$LO.
The dashed magenta line indicates the physical W boson mass, $Q=m_W$.
}
\end{figure*}

Figure~\ref{fig:PrescriptionComparison} shows the ratio $R_{W^+W^-}$ as a function of the virtuality $Q$.
The bands were created using Prescription A (left), B (middle) and C (right). 
Just like for on-shell production, we observe that correlated and uncorrelated scales lead to vastly different estimates for the size of the bands. In particular, Prescription B gives an extremely conservative estimate of the bands over the whole range of $Q$ considered. 
Unlike for on-shell production, the bands obtained from Prescription A do not overlap for $Q>200\,\textrm{GeV}$, which indicates that Prescription A does not correctly capture the size of higher-order corrections in this range of virtualities.
In general, this calls into question whether correlated scale variations should be used as an uncertainty estimator for ratios of cross sections with very similar K-factors.
Instead, Prescription C seems to give the most reliable estimator of the size of the residual perturbative corrections for ratios of cross sections over the whole range of virtualities considered. 

\begin{figure*}[!h]
\centering
\includegraphics[width=0.32\textwidth]{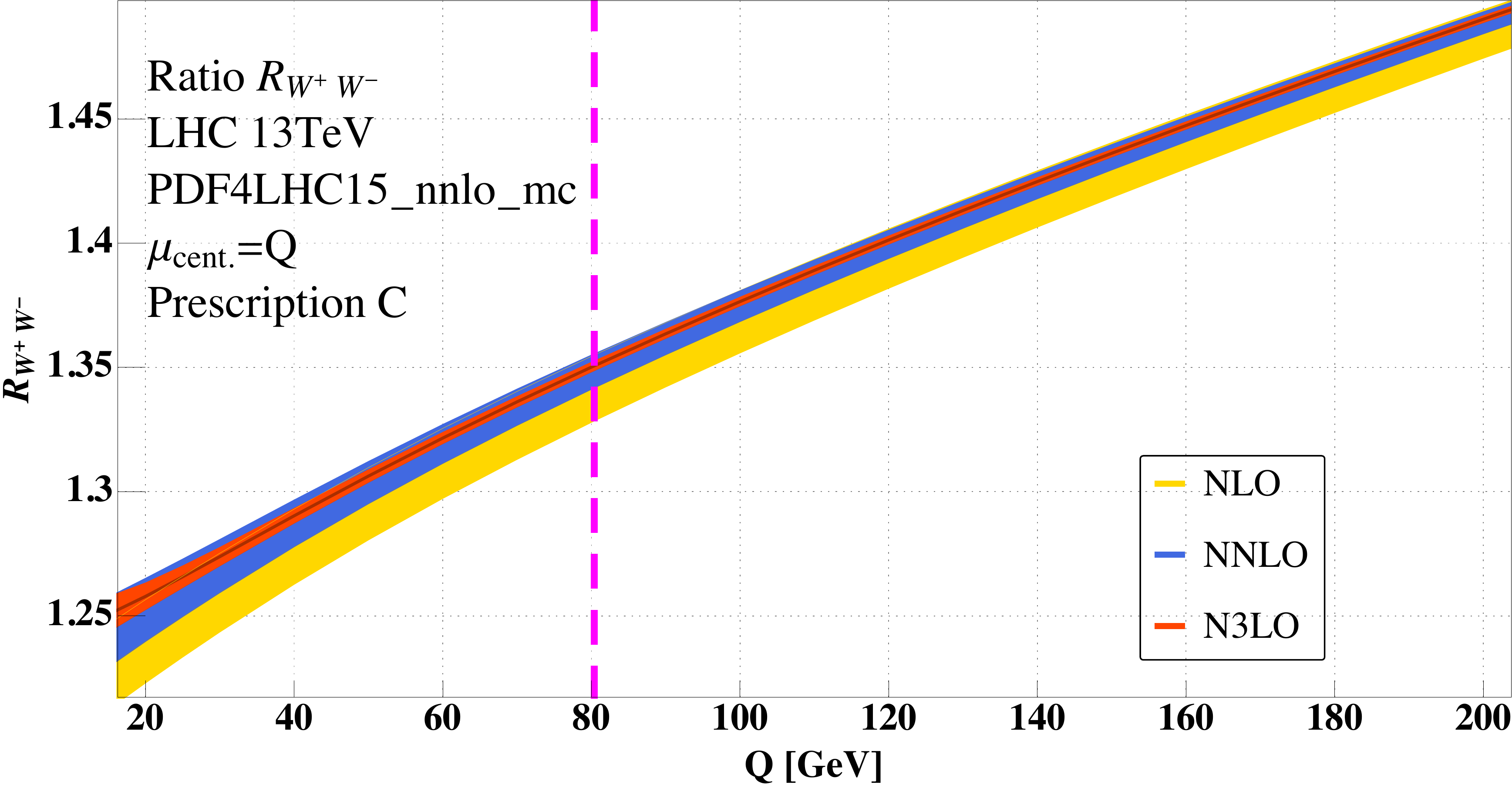}
\includegraphics[width=0.32\textwidth]{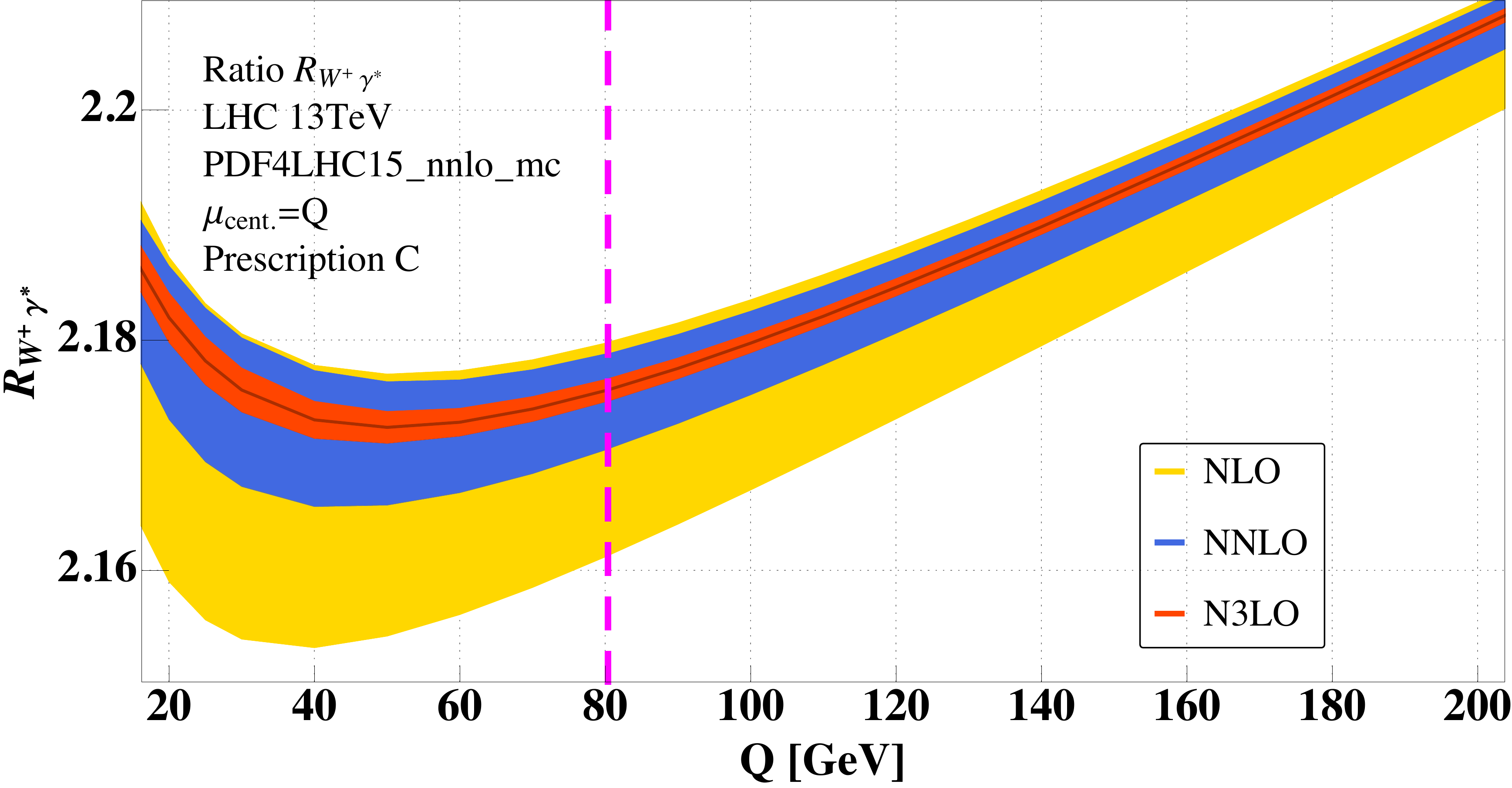}
\includegraphics[width=0.32\textwidth]{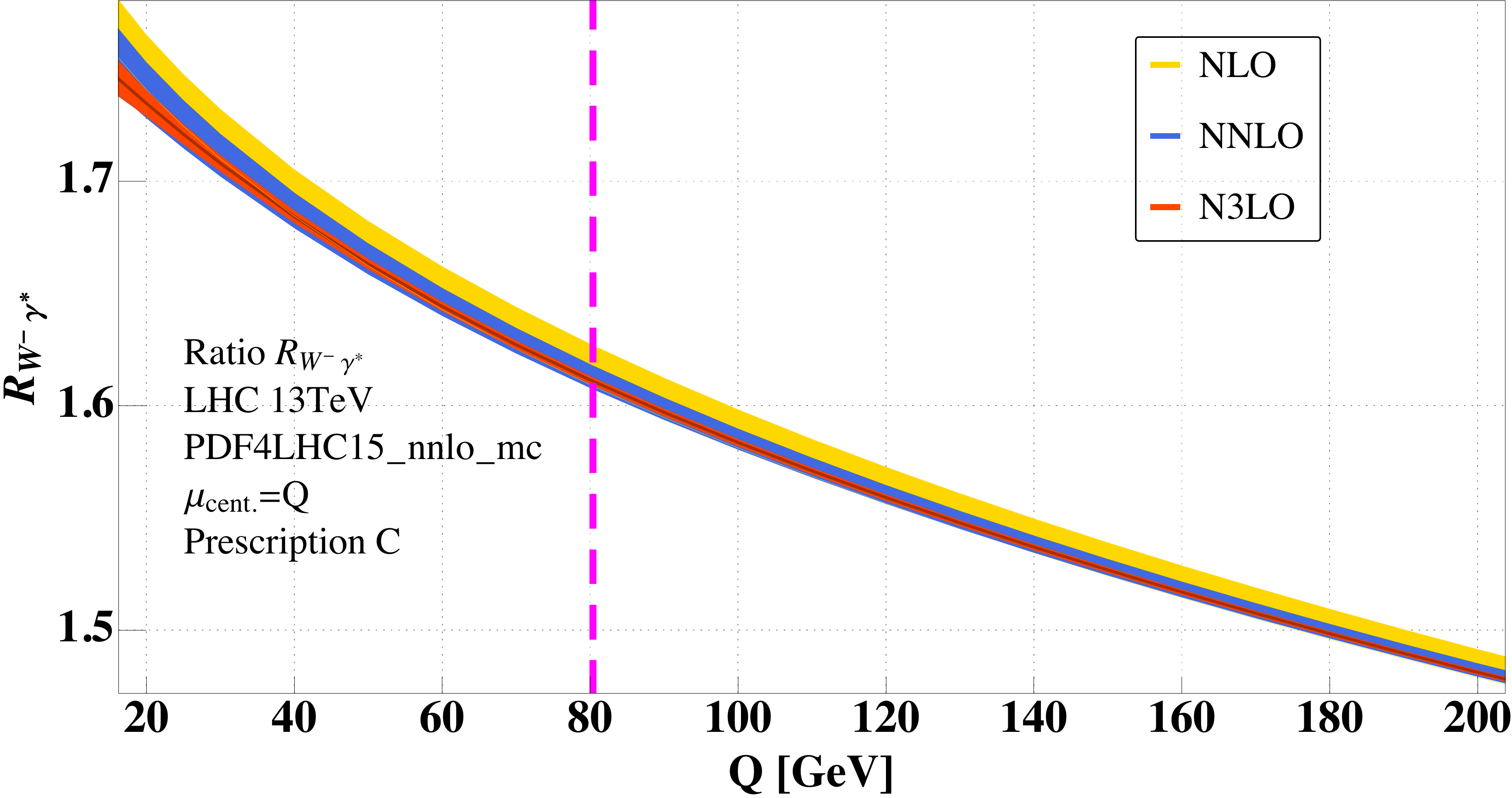}
\caption{\label{fig:Q_variation_ratio_Q}
The ratios $R_{W^+W^-}$ (left), $R_{W^+\gamma^\ast}$ (middle) and $R_{W^+\gamma^\ast}$ (right) as a function of the virtuablity $Q$. The uncertainty bands are obtained with Prescription C for the central scale $\mu^{\textrm{cent}}=Q$. 
The dashed magenta line indicates the physical W boson mass, $Q=m_W$.
}
\end{figure*}
In fig.~\ref{fig:Q_variation_ratio_Q} we extend our analysis from $R_{W^+W^-}$ (left) to $R_{W^+\gamma^\ast}$ (middle) and $R_{W^+\gamma^\ast}$ (right). In all cases the bands are estimated using Prescription C. Similar to our discussion in the previous paragraph, we find that Prescription C delivers reliable estimates also for the latter two ratios. In all cases we find that the residual perturbative uncertainty is very small, making ratios of production cross sections of electroweak gauge bosons very stable under perturbative corrections, and therefore ideal precision observables.

While Prescription C seems to give reliable estimates for all cross section ratios considered over a wide range of kinematics,
we have to point out that Prescription C has the obvious shortcoming that it gives a vanishing result whenever two consecutive perturbative orders give identical numerical predictions. 
The same fact leads to rather unconventional shapes of the uncertainty band as a function of $Q$.
Consequently, prescription C on its own would not serve as a good estimator of perturbative uncertainties. 
For example, it predicts vanishing perturbative uncertainty of $R_{W^+W^-}$ at NLO around $Q=700$ GeV, which is clearly unreliable.
In order to estimate perturbative uncertainties we therefore suggest that multiple different perspectives and estimators should be considered, in order to test the quality of the estimates obtained. For the future, it would be interesting to investigate other prescriptions to estimate the impact of missing perturbative orders, including prescriptions based on statistical methods~\cite{Cacciari:2011ze,Bonvini:2020xeo}. A more detailed study of these effects, however, would go beyond the scope of this paper.

We finish by presenting results for an observable which is closely related to the ratio $R_{W^+W^-}$. The lepton-charge asymmetry is defined as:
\beq
A_W(Q) = \frac{\sigma_{W^+}(Q) - \sigma_{W^-}(Q)}{\sigma_{W^+}(Q) + \sigma_{W^-}(Q)} = \frac{R_{W^+W^-}(Q)-1}{R_{W^+W^-}(Q)+1}\,.
\eeq

\begin{figure*}[!h]
\centering
\includegraphics[width=0.8\textwidth]{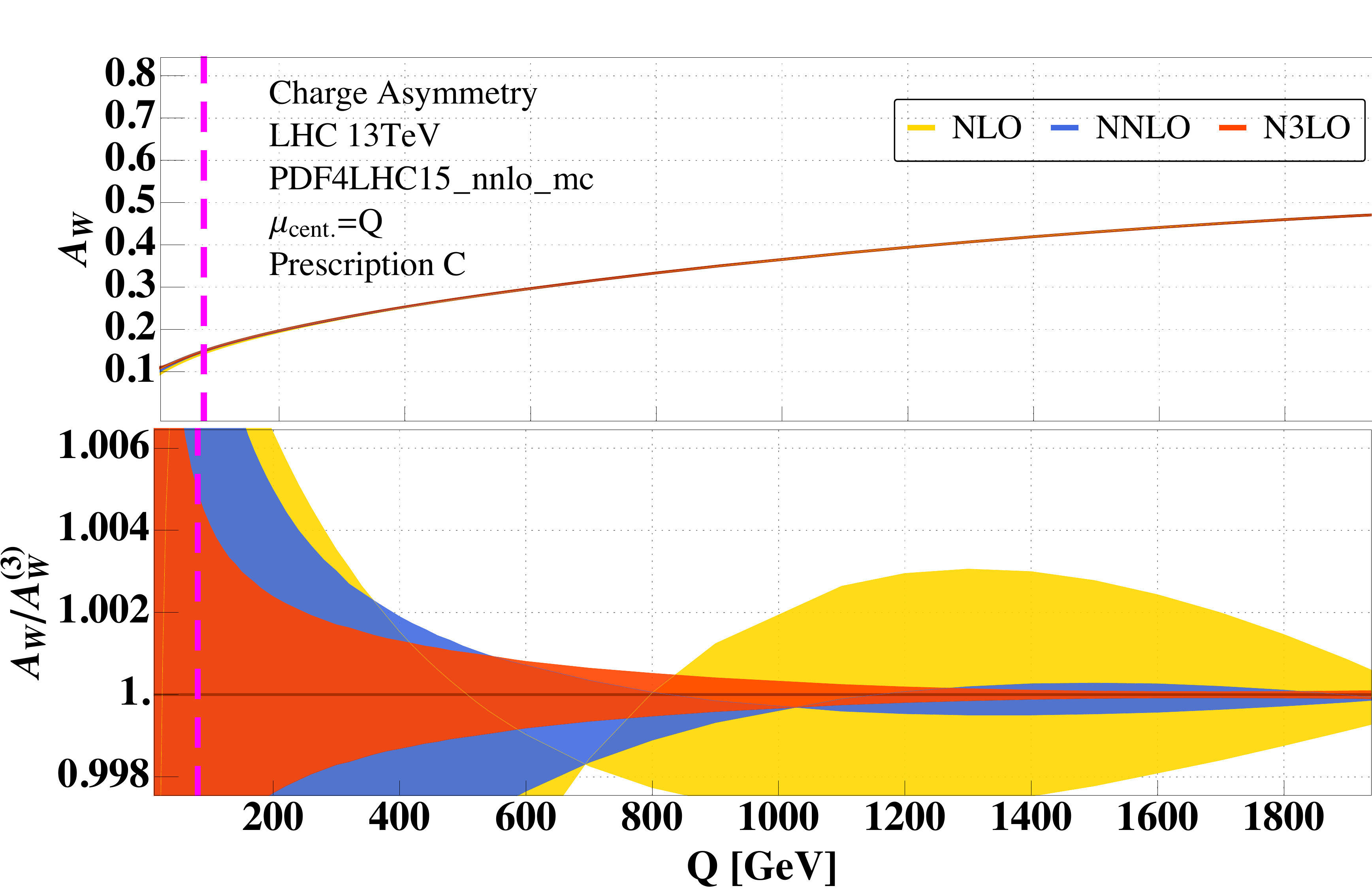}
\caption{\label{fig:AW_LQ}
The lepton-charge asymmetry $A_W$ as a function of the virtuality $Q$. The uncertainty bands are obtained with Prescription C.
The dashed magenta line indicates the physical W boson mass, $Q=m_W$.
}
\end{figure*}
Figure~\ref{fig:AW_LQ} shows our predictions for the lepton-charge asymmetry as a function of $Q$ at different orders in perturbation theory. All uncertainty bands are obtained from Prescription C.
Just like the cross section ratios studied earlier, we observe a good perturbative stability and a very small residual dependence on the perturbative scales at N$^3$LO. 
In particular, for $Q=m_W$, we find
\beq\bsp
A_W^{(2)}(m_W) &\,= 0.148_{-0.34\%}^{+0.41\%}\,,\\
A_W^{(3)}(m_W)&\, = 0.149_{-0.19\%}^{+0.15\%}\,.
\esp\eeq

So far we have only discussed the uncertainties on cross section ratios from the truncation of perturbative orders. We therefore conclude by commenting on uncertainties on ratios related to PDFs, similar to those considered in Section~\ref{sec:pheno}.
Just like in the case of the perturbative uncertainty, a choice has to be made whether or not to treat PDF and strong coupling constant variation as correlated in numerator and denominator of the ratio. 
The fact that PDFs and $\alpha_S$ are universal quantities suggests indeed a correlated treatment.
In this case, ratios of cross sections could provide a remarkable tool to reduce some of the largest theoretical uncertainties afflicting the observables in question.
However,  in neutral current and charged current DY production cross section different combinations of parton distribution functions play a dominant role. 
This could indeed spoil a correlated treatment of PDF uncertainties and should be studied in more detail.


\section{Conclusion}
\label{sec:conclusion}

In this paper we have computed for the first time the N$^3$LO corrections to the inclusive production cross section of a lepton-neutrino pair at a proton-proton collider in QCD perturbation theory. 
One of the main results of this article are analytic formul\ae\ for the partonic coefficient functions for this cross section, which we make available as ancillary material with the arXiv submission of this paper.

We have studied the phenomenological impact of our results by providing numerical results for (off-shell) $W$-production at the LHC. All our results are differential in the virtuality of the $W$ boson, or equivalently in the invariant mass of the lepton-neutrino pair. 
We find that the N$^3$LO corrections are at the level of a few percent and they stabilise the perturbative progression of the series. We have also studied ratios of cross sections for $W^+$, $W^-$ and $\gamma^*$, as well as the charge asymmetry at the LHC, i.e., the amount of produced $W^+$ bosons relative to the amount $W^-$ bosons.
We find that these ratios feature a remarkable perturbative stability and that they can be predicted with very high precision. 
For the future, it would be interesting to study, in addition to the QCD corrections discussed here, how the inclusive production probability of weak bosons is impacted by electroweak and QED corrections, for example the electroweak and mixed QCD-electroweak corrections to Drell-Yan processes of refs.~\cite{Balossini:2009sa,Denner:2009gj,Denner:2012ts,Kara:2013dua,Kilgore:2011pa,Kotikov:2007vr,Dittmaier:2014qza,Buccioni:2020cfi,Delto:2019ewv,Bonciani:2019nuy,1806554}. 
While such a study is beyond the scope of this article, we would like to stress their importance here.

Combined with the results for virtual photon production of ref.~\cite{Duhr:2020seh}, our results have also allowed us to investigate the progression of the perturbative series through N$^3$LO in QCD for one of the simplest classes of hadron collider observables, namely fully inclusive vector boson production cross sections. Understanding the perturbative convergence of this class of processes is an important proxy to understand the precision that can be reached more generally for LHC observables at this order in perturbation theory. Here we summarise our main observations and conclusions, and the possible implications for a precision physics program at the LHC. First, we observe in all cases that the N$^3$LO corrections shift the central value of the predictions by a few percent. This implies that, very likely, also for more differential observables percent-level precision can only be achieved after the inclusion of N$^3$LO corrections. It would be interesting to develop and extend techniques to perform differential calculations at N$^3$LO (see, e.g.,~refs.~\cite{Dulat:2018bfe,Dulat:2017prg,Dreyer:2016oyx,Dreyer:2018qbw,Cieri:2018oms,Billis:2019vxg} for first steps in this direction). Second, we observe that the uncertainties related to PDFs generically dominate over the residual perturbative uncertainty at N$^3$LO. This motivates to push for determining PDFs at this order in perturbation theory, by extracting them from experimental measurements confronted to theory calculations performed at the same order, and to evolve them using the (yet unknown) DGLAP evolution kernels at N$^4$LO. Finally, we observe that the conventional method to estimate the missing higher-order terms by varying the factorisation and renormalisation scales by a factor of two around a hard scale does not give reliable results at N$^3$LO. This calls for an improved method to estimate the missing higher-order terms, e.g., by studying the progression of the perturbative series as done here, or by considering statistically-motivated techniques such as those of refs.~\cite{Cacciari:2011ze,Bonvini:2020xeo}. However, we have also observed that the determination of the residual perturbative uncertainty may not be completely decoupled from the study of PDF effects: Indeed, we observe substantial cancellations between different partonic channels.  It is tantalising to speculate if these cancellations are responsible for the non-overlapping scale variation bands at N$^3$LO, and if they persist once a complete set of N$^3$LO PDFs is available. A detailed study of these effects, however, goes beyond the scope of this paper and is left for future work.

\section*{Acknowledgements}
We would like to thank L. Harland-Lang and R. Thorne for useful comments and for pointing out a mistake in fig. 8 of the first version of our article on the arXiv.
This work is supported in part by the European Research Council grant No 637019 ``MathAm'' (CD).
FD and BM were supported by the Department of Energy, Contract DE-AC02-76SF00515.
BM was also supported by the Pappalardo Fellowship.

\bibliography{Bib}
\bibliographystyle{JHEP}
\end{document}